\documentclass[12pt]{article}
\usepackage{graphicx}
\usepackage{amssymb}
\usepackage{amsthm}
\usepackage{lineno}
\usepackage{color}
\usepackage{float}
\usepackage{tikz}
\usepackage{amsmath}
\usepackage{booktabs,subcaption,amsfonts,dcolumn}
\usepackage{multirow}
\usepackage[graphicx]{realboxes}
\usepackage{arydshln}

\usepackage{mathtools}
\usepackage{multirow,tabularx}
\usepackage{adjustbox}
\usepackage[colorlinks,citecolor=blue,urlcolor=blue]{hyperref}
\usepackage{marginnote}
\usepackage{times}
\usepackage{natbib}
\usepackage{tcolorbox}

\newcommand{\cond}{\,|\,}

\begin{document}

\title{Efficiency gain in association studies based on population surveys by augmenting outcome data from the target population}

\author{Tommi H\"ark\"anen$^a$, Sangita Kulathinal$^b$, Arya Panthalanickal Vijayakumar$^c$}
\date{%
    $^a$ Finnish Institute for Health and Welfare, Helsinki, Finland \\
    $^b$ Department of Mathematics and Statistics, University of Helsinki, Helsinki, Finland\\
$^c$ Department of Community Medicine, UiT The Arctic University of Norway, Tromsø, Norway\\[2ex]%
    \today \\
    \vspace{3cm}
Word count - 6492 \\
\vspace{1cm}
Funding - NA
}
\maketitle

\newpage

\begin{abstract}
Routinely collected nation-wide registers contain socio-economic and health-related information from a large number of individuals. However, important information on lifestyle, biological and other risk factors is available at most for small samples of the population through surveys. A majority of health surveys lack detailed medical information necessary for assessing the disease burden. Hence, traditionally data from the registers and the surveys are combined to have necessary information for the survey sample. Our idea is to base analyses on a combined sample obtained by adding a (large) sample of individuals from the population to the survey sample.
The main objective is to assess the bias and gain in efficiency of such combined analyses with a binary or time-to-event outcome.
We employ  
(i) the complete-case analysis (CCA) using the respondents of the survey, 
(ii) analysis of the full survey sample with both unit- and item-nonresponse under the missing at random (MAR) assumption and (iii) analysis of the combined sample under mixed type of missing data mechanism.
We handle the missing data using multiple imputation (MI)-based analysis in (ii) and (iii). We utilize simulated as well as empirical data on ischemic heart disease obtained from the Finnish population.
Our results suggested that the MI methods improved the efficiency of the estimates when we used the combined  data for a binary outcome, but in the case of a time-to-event outcome the CCA was at least as good as the MI using the larger datasets, in terms of the the mean absolute and squared errors. 
Increasing the participation in the surveys and having good statistical methods for handling missing  covariate data when the outcome is time-to-event would be needed for implementation of the proposed ideas.

\end{abstract}

\pagebreak

\begin{center}
\fbox{
    \parbox{0.9\textwidth}{
        \textbf{Statement of Significance}
        Collecting data for various lifestyle-related risk factors through surveys is generally expensive, whereas various socio-demographic and health outcome data from population registers are readily available and therefore inexpensive to obtain for research purposes.
        We aim to combine a relatively small survey sample with a large sample from the target population drawn from registers. We then utilize statistical methods developed for handling missing data to possibly carry out better estimation of association between risk factors and health outcomes.  For a binary outcome, the standard statistical methods improved the accuracy, while for the time-to-event outcomes, e.g. disease onset or death, better methods are needed.

    }
}
\end{center}
\section{Introduction}

\label{S:1}

Collecting data for epidemiological studies from population registers is often a cost-effective way of obtaining large datasets for statistical analyses. However, information on all risk factors of interest are not available in such registers, and expensive population surveys are needed in addition to the  register data. Commonly, statistical analyses are based on the survey data merged with the register data of the survey participants, which does not change the number of outcome events and hence, the statistical power of association studies may reduce. A potential improvement could be obtained by selecting a large (random) sample of individuals from the target population using the population registers, and using survey, comparatively much smaller, as the main source of the risk factors data. This approach, though attractive as the missingness mechanism can be controlled by the researcher to some extent, creates a large proportion of missing risk factors data requiring careful choice of statistical methods for the analyses. We consider an integrated population-based health examination survey and register-based study, wherein a part of the data are obtained by health examination surveys based on relatively small sample sizes, while demographic data and disease outcomes (disease status and time to the disease), are retrieved from  the nation-wide population and health registers.

The registers generally have good coverage but many important risk factors can only be obtained through surveys/examinations. A major issue in health examination surveys is the decreasing response rates (\cite{little_rubin_2019}). Every participant may not undergo the full extent of examinations or respond to all queries. The tendency for non-participation is frequently linked to adverse health outcomes, leading to a scenario where morbidity and mortality rates are discernibly higher among non-participants \citep{TolonenDobsonEtAlEte2005,harald2007non,AlkerwiSauvageotEtAlCpa2010,little_rubin_2019}. It is hypothesized that  the  risk factors are distributed differently between the non-participants and participants, and that in turn, increase incidences of adverse health outcomes in non-participants compared to participants. To summarise, the issue of missing data primarily affects the variables collected through questionnaires or direct measurements within the survey, and not the register-based data. 

A variety of statistical approaches have been developed to address the issue of missing data \citep{MolenberghsKenward_2007}. The idea of Multiple Imputation (MI), as means of statistical inference, was introduced in \citep{rubin1996multiple}, which has since gained prominence for its effectiveness in managing missing data. The MI is particularly advantageous for end-users, who typically only have access to the dataset compiled by the database constructors \citep{rubin1996multiple}. The core principle of MI involves replacing each missing entry with multiple plausible values, derived from a predictive distribution based on the observed data. MI excels by accounting for the full spectrum of variability and uncertainty inherent in the data providing a robust framework for deriving valid conclusions from datasets marred by incompleteness. In the context of the Health 2000 Survey, prevalence estimates based on MI were found more accurate compared to alternative methods \citep{HeistaroMrH2008, harkanen2016systematic}. On the other hand, more complex distributions, for example in time-to-event analyses, pose special challenges in statistical modeling and therefore also in selecting appropriate MI methods.

In the field of prognostic research, the Multivariate Imputation by Chained Equations (MICE), also known as Fully Conditional Specification (FCS), is currently regarded as the golden standard \citep{van2018flexible}.  MICE is a special MI technique in which, multivariate missing data are imputed iteratively on a variable-by-variable basis. It approximates the multivariate distribution through a set of conditional densities, in which each variable with missing values is modeled using all other variables. This method is particularly adept at handling variables of diverse measurement scales, making it invaluable for handling survey data predominantly involved in collecting categorical data, and non-monotone missing data patterns. The challenge of imputing time-to-event data has been addressed by White and Royston \citep{white2009imputing}, who proposed using Nelson-Aalen estimates of cumulative hazard and event status as covariates in the imputation model.

The primary objective of this paper is to evaluate potential efficiency gains contrasted with increasing bias, when integrating observed survey data with an additional, large sample of individuals from population registers (Figure \ref{fig:sub1a}), and applying multiple imputation for risk factor variables that were only collected in the survey but were missing for individuals who did not participate or were in the additional sample. This approach is contrasted with traditional methods that either depend exclusively on complete-case data, or utilize survey data augmented with population register variables solely for individuals who participated in the survey, with MI applied to address the item-nonresponse within this limited dataset.
In Section~\ref{S:2}, we outline the health examination survey \citep{HeistaroMrH2008}, detailing the collection of various risk factors and the subsequent follow-up on ischemic heart disease (IHD) through health registers. Section~\ref{sec::methods} delves deeper into the methodologies employed for handling  missing data. 

We execute three distinct analyses, utilizing different subsets of data:
\begin{enumerate}
    \item a complete case analysis (CCA) from the survey, which has been found inadequate in many applications, based on the survey,
    \item all survey participants, including instances of missing data, and
    \item combined survey and a large population sample data.
\end{enumerate}
In scenarios devoid of missing survey data, the first and second options are the same. For the second and third analytical approaches, MI techniques are utilized to address missing data concerns. Section~\ref{sec:simulation} presents simulation results comparing MI and CCA methods, focusing on binary and time-to-event outcome data separately. Our work is motivated by the integrated population-based association study of ischemic health disease and its risk factors where the data from the Health 2000 survey and the Finnish population and hospital discharge registers are combined.  

\section{Empirical study on Ischemic Heart Disease (IHD) data}
\label{S:2}

Our empirical study on IHD consists of the Health 2000 Survey combined with the Finnish population and the Care Register for Health Care \citep[Hilmo;][]{hilmo}. The association analysis is based on the demographic ($Z$), risk factors ($X$) and disease outcome ($Y$) data.

\subsection{The Health 2000 Survey}
\label{S:2-H2000}

The Health 2000 Survey was a Finnish national health examination survey carried out in year 2000-2001 \citep{HeistaroMrH2008}. A sample was selected from the population using the Finnish population register on July 1, 2000 (baseline), of which 8028 were in the age group of 30 years and above, and 7938 were found to be free from IHD at the baseline, referred to as the cohort $\mathcal{C}$ of healthy individuals below.

The risk factors recorded in the survey were total and HDL cholesterol, two measures of systolic and diastolic blood pressure, smoking habits (with categories doesn't smoke and smoke), body mass index (BMI, $\text{kg}/m^2$), alcohol usage (abstainer whole life; used earlier but not now; and used earlier and now), overuse of alcohol (no; yes with $\ge4$ for women and $\ge7$ for men), past year alcohol consumption frequency (no; often; and seldom) and leisure time physical activity (no physical activity or stress; walking, biking etc. 4 hrs/week; exercise at least 3 hrs/week and exercise many times a week). The descriptive summary of the risk factors is given in Table~\ref{tab: covar}. Out of the healthy individuals, 4792 individuals, denoted as $\mathcal{R}$, had fully observed analysis variables.
The cohort was followed up till June 30, 2018 for the first hospitalization due to IHD based on the ICD-10 codes I20-I25 or death. The time of the hospitalization or death were obtained  from the Care Register for Health Care \citep[Hilmo;][]{hilmo} or from the population register. The survey sample was linked with the registers using the unique personal identification codes.

\begin{figure}[htb]
\centering
    \begin{subfigure}[t]{0.45\textwidth}
        \tikzset{every picture/.style={line width=0.75pt}}         
    \begin{tikzpicture}[x=0.75pt,y=0.75pt,yscale=-1,xscale=1]
\draw   (12,11.5) -- (283,11.5) -- (283,236.5) -- (12,236.5) -- cycle ;
\draw  [fill={rgb, 255:red, 74; green, 144; blue, 226 }  ,fill opacity=0.45 ] (21,54.9) .. controls (21,49.43) and (25.43,45) .. (30.9,45) -- (263.1,45) .. controls (268.57,45) and (273,49.43) .. (273,54.9) -- (273,84.6) .. controls (273,90.07) and (268.57,94.5) .. (263.1,94.5) -- (30.9,94.5) .. controls (25.43,94.5) and (21,90.07) .. (21,84.6) -- cycle ;
\draw  [fill={rgb, 255:red, 176; green, 4; blue, 23 }  ,fill opacity=0.45 ] (21,138.3) .. controls (21,127.36) and (29.86,118.5) .. (40.8,118.5) -- (258.2,118.5) .. controls (269.14,118.5) and (278,127.36) .. (278,138.3) -- (278,197.7) .. controls (278,208.64) and (269.14,217.5) .. (258.2,217.5) -- (40.8,217.5) .. controls (29.86,217.5) and (21,208.64) .. (21,197.7) -- cycle ;
\draw  [fill={rgb, 255:red, 13; green, 251; blue, 167 }  ,fill opacity=0.62 ] (27.7,167) -- (271.3,167) -- (271.3,212.5) -- (27.7,212.5) -- cycle ;

\draw (27,18.5) node [anchor=north west][inner sep=0.75pt]   [align=left] {{\large {\fontfamily{ptm}\selectfont Population}}};
\draw (32.36,49) node [anchor=north west][inner sep=0.75pt]   [align=left] {{\fontfamily{ptm}\selectfont Random sample of the population}\\$\displaystyle ( z_{i} ,y_{i}) ,\ i\in \ P$};
\draw (44.92,124) node [anchor=north west][inner sep=0.75pt]   [align=left] {{\fontfamily{ptm}\selectfont Health examination survey}\\$\displaystyle ( z_{i} ,y_{i}) ,\ i\in \ C$};
\draw (45.15,178) node [anchor=north west][inner sep=0.75pt]   [align=left] {{\fontfamily{ptm}\selectfont Participants} $\displaystyle ( z_{i} ,x_{i} ,y_{i}) ,\ i\in \ R$};
\end{tikzpicture}
        \caption{Observed data in the three subsets of the analysis data.}
        \label{fig:sub1a}
    \end{subfigure}
    \hfill 
    \begin{subfigure}[t]{0.45\textwidth}
        \tikzset{every picture/.style={line width=0.75pt}}       
        \begin{tikzpicture}[x=0.75pt,y=0.75pt,yscale=-1,xscale=1]

\draw [color={rgb, 255:red, 74; green, 74; blue, 74 }  ,draw opacity=1 ][line width=1.5]    (10,228.57) -- (279,229.49) ;
\draw [shift={(282,229.5)}, rotate = 180.2] [color={rgb, 255:red, 74; green, 74; blue, 74 }  ,draw opacity=1 ][line width=1.5]    (14.21,-4.28) .. controls (9.04,-1.82) and (4.3,-0.39) .. (0,0) .. controls (4.3,0.39) and (9.04,1.82) .. (14.21,4.28)   ;
\draw  [dash pattern={on 4.5pt off 4.5pt}]  (25.17,25) -- (25.17,232.1) ;
\draw  [dash pattern={on 4.5pt off 4.5pt}]  (254.91,25) -- (254.91,232.1) ;
\draw [color={rgb, 255:red, 128; green, 128; blue, 128 }  ,draw opacity=1 ][line width=1.5]    (26.25,183.22) -- (252.64,183.68) ;
\draw [shift={(255.99,183.69)}, rotate = 0.12] [color={rgb, 255:red, 128; green, 128; blue, 128 }  ,draw opacity=1 ][line width=1.5]      (0, 0) circle [x radius= 4.36, y radius= 4.36]   ;
\draw [color={rgb, 255:red, 128; green, 128; blue, 128 }  ,draw opacity=1 ][fill={rgb, 255:red, 80; green, 227; blue, 194 }  ,fill opacity=1 ][line width=1.5]    (26.25,125.72) -- (173.63,126.65) ;
\draw  [color={rgb, 255:red, 74; green, 144; blue, 226 }  ,draw opacity=1 ][fill={rgb, 255:red, 74; green, 144; blue, 226 }  ,fill opacity=1 ][line width=1.5]  (174.72,127) .. controls (174.72,124.61) and (176.96,122.68) .. (179.73,122.68) .. controls (182.5,122.68) and (184.74,124.61) .. (184.74,127) .. controls (184.74,129.39) and (182.5,131.33) .. (179.73,131.33) .. controls (176.96,131.33) and (174.72,129.39) .. (174.72,127) -- cycle ;
 
\draw [color={rgb, 255:red, 128; green, 128; blue, 128 }  ,draw opacity=1 ][line width=1.5]    (26.25,69.62) -- (119.45,69.62) -- (222.4,69.62) ;
 
\draw  [color={rgb, 255:red, 208; green, 2; blue, 27 }  ,draw opacity=1 ][fill={rgb, 255:red, 208; green, 2; blue, 27 }  ,fill opacity=1 ] (223.62,63.89) -- (235.54,63.89) -- (235.54,74.18) -- (223.62,74.18) -- cycle ;

\draw (206,42.39) node [anchor=north west][inner sep=0.75pt]   [align=left] {{\fontfamily{ptm}\selectfont Death}};
 
\draw (167,104.1) node [anchor=north west][inner sep=0.75pt]   [align=left] {{\fontfamily{ptm}\selectfont IHD}};
 
\draw (236,158.32) node [anchor=north west][inner sep=0.75pt]   [align=left] {{\fontfamily{ptm}\selectfont Censored}};
 
\draw (85,239) node [anchor=north west][inner sep=0.75pt]   [align=left] {{\fontfamily{ptm}\selectfont Calendar time}};
 
\draw (-2,236) node [anchor=north west][inner sep=0.75pt]  [font=\small] [align=left] {01/07/2000};
 
\draw (221,233) node [anchor=north west][inner sep=0.75pt]   [align=left] {{\small 30/06/2018}};

\end{tikzpicture}
        \caption{Time-to-event framework for  follow-up of IHD or death.} 
        \label{fig:sub1b}
    \end{subfigure}
 
\caption{Structure of the samples and the follow-up data.  (a) Background register data $z_i$ including age, sex, place of residence and education, and outcomes $y_i$ are available for the random sample of the  population $\mathcal{P}$ and survey sample $\mathcal{C}$. Risk factors $x_{1i}$ are available for $i\in\mathcal{C}$ while $(x_{1i},\,x_{2i})$ are available only for survey participants $\mathcal{R}$.}
\label{fig:IHD_data_rep}
\end{figure}
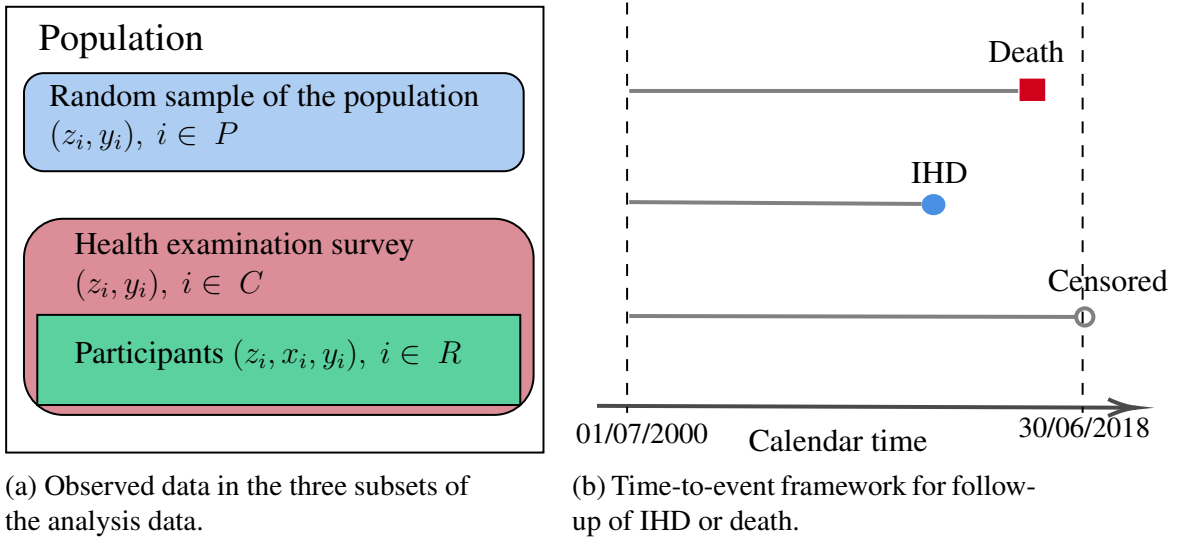
\subsection{Augmenting the cohort by a random sample from the population}
\label{S:2-reg}
Finnish population and healthcare registers are renowned for their comprehensive coverage \citep{tolppanen2013use}, and therefore register-based analyses can improve the efficiency of estimates of population health indicators~\citep{harald2007non,harkanen2016systematic}. The survey cohort ($\mathcal{C}$) was sampled from the population register, which, as of July 1, 2000, included detailed information on age and sex. Furthermore, the population register and Hilmo Register offered the capability to monitor all individuals aged 30 and above, from July 1, 2000, to June 30, 2018, for incidences of IHD and mortality. As a result, we gathered an additional random sample comprising 484,483 individuals, which is about 15\% of the target population of the survey, as of July 1, 2000; this additional sample is denoted as $\mathcal{P}$. The risk factor data for the sample in $\mathcal{P}$ are missing because they were not part of the original survey. 
However, we were able to document their background information and outcomes, specifically regarding IHD and mortality. Figure~\ref{fig:sub1a} illustrates the comprehensive data set including the population~($\mathcal{P}$), cohort~($\mathcal{C}$), and full data~($\mathcal{P}\cup\mathcal{C}$), whereas Figure~\ref{fig:sub1b} delineates the potential trajectories during the follow-up period for IHD or mortality. The associated risk factors, amounting to a total of 16 variables, along with the time-to-event data, are detailed in Table~\ref{tab: covar}. 

Individuals not included in the survey, i.e. individuals in~$\mathcal{P}$, lacked corresponding risk factor data, which were otherwise collected during the survey. Consequently, merging the population register data with the data from the Health 2000 Survey, both following the same timeline of the study and including these crucial risk factors results in a significant volume of missing risk factor information, as illustrated in Figure~\ref{fig:sub1a}.

\begin{table}[htbp]
\centering
\caption{Descriptive statistics of the integrated data from the Health 2000 survey sample ($\mathcal{C}$, sample size = 7938) and register data ($\mathcal{P}\cup\mathcal{C}$, sample size in $\mathcal{P}$ = 484,483). $N$ = number of observations. }
 
\label{tab: covar}
\renewcommand{\arraystretch}{1.5}
\begin{subtable}{\textwidth}
\centering
\caption{Continuous factors. IQR is the interquartile range.}
\label{tab:cont}
\scalebox{0.725}{
\begin{tabular}{lcccccl}
\toprule
\multirow{2}{*}{Covariates} & \multirow{2}{*}{\begin{tabular}[c]{@{}l@{}}Observed\\ values (N)\end{tabular}} & \multicolumn{2}{c}{Missingness ($\%$) in} & \multirow{2}{*}{Mean $\pm$ SD} & \multirow{2}{*}{Range} & \multirow{2}{*}{\begin{tabular}[c]{@{}l@{}}Median \\  (IQR)\end{tabular}} \\
&  & \begin{tabular}[c]{@{}l@{}}$\mathcal{P}\cup\mathcal{C}$\end{tabular} & \begin{tabular}[c]{@{}l@{}}$\mathcal{C}$\end{tabular} &  &  &  \\ \hline
Age & 492,421 & 0.0 & 0.0 & 63 $\pm$ 18.5 & (30--117) & 61 (50--76) \\
\hline

Total cholesterol & 6626 & 98.7 & 16.5 & 6.0 $\pm$ 1.11 & (1.9--11.7) & 5.9 (5.2--6.6) \\
HDL cholesterol & 6626 & 98.7 & 16.5 & 1.3 $\pm$ 0.38 & (0.2--3.4) & 1.2 (1--1.5) \\
\hdashline
Systolic BP1 & 6677 & 98.6 & 15.9 & 139.6 $\pm$ 22.41 & (68--245) & 138 (124--154) \\
Diastolic BP1 & 6677 & 98.6 & 15.9 & 81.8 $\pm$ 12.21 & (0--134) & 82 (74--90) \\
Systolic BP2 & 6654 & 98.6 & 16.2 & 137.6 $\pm$ 22.08 & (66--236) & 136 (122--150) \\
Diastolic BP2 & 6652 & 98.6 & 16.2 & 80.8 $\pm$ 11.53 & (0--131) & 80 (74--88) \\
\hdashline 
BMI & 7163 & 98.5 & 9.8 & 26.8 $\pm$ 4.68 & (12.1--53.7) & 26.2 (23.5--29.4) \\
\hdashline
\begin{tabular}[c]{@{}l@{}} Previous year \\ \ alcohol consumption\end{tabular} & 4990 & 99.0 & 37.1 & 7.5 $\pm$ 5.995 & (1--50) & 7 (2--12) \\
Total alcohol g/year & 6582 & 98.7 & 17.1 & 3833 $\pm$ 8940 & (0.229--394) & 633.5 (0--3708) \\ \hline
\end{tabular}
}
\end{subtable}
\\[1em]
\begin{subtable}{\textwidth}
\centering
\caption{Categorical factors. $*$: category "no" is not shown. Alcohol overuse for men $>$ 7 units and women $>$ 4 units.}
\vspace{-1em}
\label{tab:cat}
\scalebox{0.725}{
 
\begin{tabular}{lrrrrlr} \\ \hline
\multirow{2}{*}{} & \multirow{2}{*}{\begin{tabular}[c]{@{}l@{}}Observed\\ values (N)\end{tabular}} & \multicolumn{3}{l}{Missingness (\%) in} & \multirow{2}{*}{Categories} & \multirow{2}{*}{\begin{tabular}[c]{@{}l@{}}Frequency\end{tabular}} \\
 &  & $\mathcal{P}\cup\mathcal{C}$ & $\mathcal{C}$ & & &  \\ \hline

Gender & 492,421 & 0.0 & 0.0 & & - male  &  235,870 \\
\hline
Regular smoking status$*$ & 7,282 & 98.5 & 8.3 & & - smoke & 1,596 \\\hdashline
\begin{tabular}[c]{@{}l@{}}Physical stress \\  and activity$*$\end{tabular} & 6,542 & 98.7 & 17.6 & & \begin{tabular}[c]{@{}l@{}} - walking, biking etc. 4 hrs/week \\ - exercise at least 3 hrs/week \\ - exercise many times a week\end{tabular} & \begin{tabular}[c]{@{}r@{}} 3,508\\  1,032 \\  82\end{tabular} \\\hdashline
Alcohol consumption$*$ & 6,410 & 98.7 & 19.2 & & \begin{tabular}[c]{@{}l@{}} - used earlier but not now \\ - used earlier and now\end{tabular} & \begin{tabular}[c]{@{}r@{}} 368 \\  4,890\end{tabular} \\\hdashline

Overuse of alcohol$*$ & 6,695 & 98.6 & 15.7 & & \begin{tabular}[c]{@{}l@{}} - yes\end{tabular} & \begin{tabular}[c]{@{}r@{}}  1,914\end{tabular} \\\hdashline
\begin{tabular}[c]{@{}l@{}} 
Past year alcohol \\ consumption frequency$*$ 
 
\end{tabular} 
& 7,026 & 98.6 & 11.5 & & \begin{tabular}[c]{@{}l@{}} - often \\ - seldom\end{tabular} & \begin{tabular}[c]{@{}r@{}} 2,567 \\  3,069\end{tabular} \\ \hline
\end{tabular}}
\end{subtable}
\end{table}

\section{Missingness pattern, mechanism and methods}\label{sec::methods}

The risk factor data $X$ are missing for the population sample taken from the register ($\mathcal{P}$) and for the non-responders of the Health 2000 survey ($\mathcal{C}\setminus \mathcal{R}$).  The missingness mechanism $R_{C}$ in the survey is very likely of the missing not at random (MNAR) type, whereas the mechanism $R_{P}$ in the population sample is missing completely at random (MCAR), as the additional sample was selected randomly from the population. Our main interest is in the conditional distribution of $Y = (T, \delta)$ given $(Z,\,X)$, where $T$ is the time of an event of interest (IHD or death whichever occurs first) or censoring time, $\delta$ is the event indicator, $Z$ is the set of demographic factors/covariates obtained from the registers with no missing values, and $X$ are the risk factors/covariates from the survey data that have missing values and needed to be imputed (Table~\ref{tab: covar}).

\subsection{Regression analyses}

To investigate associations between risk factors and outcomes, either binary or time-to-event, we employ conventional logistic and survival regression models as outlined in \cite{clayton1993statistical}. For both models, we carry out three sets of analyses using: (i) using complete case data $\mathcal{R}$ from the survey, which encompasses records without any missing values (Complete Case Analysis, CCA); (ii) utilizing all available cases from the survey sample $\mathcal{C}$; and (iii) integrating the sample of the population data $\mathcal{P}$, which lacks observed values for the risk factors, with the survey sample $\mathcal{P}\cup\mathcal{C}$. We regard the analyses of (ii) as the benchmark for comparison due to its comprehensive inclusion of available data, contrary to the preliminary assumption that CCA might serve as the standard. The results obtained from approaches (i) and (iii) are then evaluated against this benchmark.

In a practical scenario to illustrate our methods, we apply the Weibull regression models to analyze transitions from a 'healthy' state to 'IHD' or 'death' within the time-to-event framework.

\subsection{Complete case analysis (CCA)}

Complete Case Analysis (CCA) refers to a statistical approach that considers only those participants without any missing values in the variables under analysis. Its popularity stems from ease of implementation and computational efficiency. CCA is deemed suitable when the mechanism of missing data is Missing Completely at Random (MCAR), as it ensures unbiased parameter estimates although missing data cause loss of statistical power. In our context, it's crucial to recognize that the missing data within the population are considered MCAR, given that both the survey sample and the population data represent random samples from the overall population, eliminating any selection bias.

However, CCA may yield biased estimates in scenarios where the missing data mechanism is Missing at Random (MAR), as the likelihood of survey participation is influenced by the observed data. Similarly, under the Missing Not at Random (MNAR) scenario, where the probability of missingness also depends on unobserved data, CCA's reliability diminishes. Although CCA can be efficient with minimal missing data,  in most other situations discarding partially observed cases may lead to a reduction in efficiency of estimation and statistical power ({\cite{MolenberghsKenward_2007,little_rubin_2019}).
\subsection{Multiple imputation}
Introduced by~\cite{rubin1978bayesian, rubin_1987}, MI has emerged as a leading method for addressing missing data, especially under the MAR assumption. MI is considered a practical approach also for sensitivity analyses where data might be missing not at random (MNAR), despite its original formulation under the MAR assumption. 
MI operates on the Bayesian principle, imputing missing entries with values usually drawn from an approximate posterior predictive distribution that is conditioned on the observed data. The core idea behind MI lies in replacing each missing entry with a set of $m\geq3$ plausible values, generating $m$ complete datasets with the same values for the non-missing part but varying values for the missing part. Each dataset is then analyzed separately with some statistical analysis of interest, and the resulting estimates and their variances are pooled for final inference.

The choice of $m$, the number of imputations, balances between the accuracy of results and computational demands. While~\cite{Schafer_1999} recommended a modest $m$, typically between 5 to 10, later studies, such as~\cite{WhiteEtAl2011Multiple}, advocated for $m$ to be at least as large as the percentage of cases with missing information. For our analysis involving the combined data $\mathcal{P}\cup\mathcal{C}$, with missing values approaching 100\% as demonstrated in Table~\ref{tab: covar}, we opted for $m=25$ to manage computational constraints effectively.

To approximate the Bayesian predictive distribution, particularly when direct analytical solutions are infeasible or the dataset size prohibits exact (or near exact) methods, methods based on Fully Conditional Specification (FCS), or Multiple Imputation by Chained Equations (MICE), offer a balanced solution \citep{van2018flexible}. FCS imputes missing data through iterative variable-by-variable imputation, drawing from full-conditional distributions based on both observed and currently imputed values of other variables \citep{van2006fully, van2007multiple}. Despite assumptions of the existence of a proper multivariate joint distribution, which can be impossible to verify, practical applications often proceed without confirming its existence, facilitated by FCS's flexibility to handle variables of different scales under the Generalized Linear Models (GLM) framework.

For selecting regression models within FCS, especially when evaluating interactions or nonlinearities, the nonparametric Classification and Regression Tree (CART) method \citep{breiman1984classification} is utilized for its innate ability to adapt to underlying complexity of the dataset.
\subsubsection{MI method for time-to-event outcome}

Dealing with time-to-event data, characterized by the event status and its timing, introduces specific challenges for MI. \cite{white2009imputing} proposed approximation of the full conditional distribution using the Nelson-Aalen estimate for cumulative hazard functions alongside the event status within the imputation model. This strategy outperformed simpler approaches using either the follow-up duration, $T$, or its logarithmic transformation, $\log(T)$. It is adept for imputing missing values in both continuous and binary covariates through linear or logistic regression models, respectively.

Time-to-event data intricacies often stem from varying time scales such as calendar time, chronological age or duration. For instance, age might better encapsulate hazard rate changes than merely follow-up time, as suggested by \cite{white2009imputing} and implemented in the latest \texttt{mice} package version in R. However, due to the delayed entry, the observations are conditional on the disease-free survival until the baseline age, thus, for example, in the Poisson likelihood the integral of the hazard function is from the baseline age to the age at the end of follow-up. To accommodate delayed entries, we adapted the R function to account for both start and stop times of follow-up, details of which are disclosed in the Appendix.

For MI of risk factors of time-to-event data, we use the CART and GLM models. 
In addition to the observed background variables obtained from registers, the predictors in the imputation models are the event indicators, and either the Nelson-Aalen estimates from the baseline age to the end of follow up, the follow-up time $T$ or its logarithm $\log(T)$. 
The covariates are selected in each of the imputation model using Kendall $\tau$ correlation value with the outcome.

\section{Simulation Study}\label{sec:simulation}
For convenience, we have assumed that the survey data are completely observed so that in this case $\mathcal{R} = \mathcal{C}$ and we will write $\mathcal{R}$ hereafter in this section.
we have excluded handling and evaluation of survey data ($\mathcal{C}$) with non-response missingness, as this issue has been sufficiently explored in the area of missing data, particularly when the rate of missingness is not excessively high. In the simulation, 

\subsection{Study Design}
The data were simulated to closely resemble the combined characteristics of the empirical data, denoted as $\mathcal{P} \bigcup \mathcal{R}$. This includes the register data ($\mathcal{P}$) which comprises variables such as age, sex  which were available for the entire sample, and from the survey data ($\mathcal{R}$) which integrated risk factor variables along with the register variables, without any missing values.  A total ($\mathcal{P} \bigcup \mathcal{R}$) of 100,000 observations were generated for the binary and the time-to-event outcomes, and missing values only for the covariates resulting in different sized $\mathcal{R}$. 

\subsubsection{Covariates}
\label{sec: sim covariates}
 
For both the analysis models, we simulated five covariates. Covariates $(X_1, X_2, X_3, X_4)$ were from the following normal distributions.
\begin{eqnarray*}
X_1  & \sim & \mathcal{N}(2,\,1^2), \;  X_2  \sim  \mathcal{N}(-2,\,1^2), \\
(X_3\cond x_1,\,x_2) & \sim & \mathcal{N}(0.5x_1 + 0.5x_2,\,2^2), \;  \text{and} \\ 
(X_4\cond x_1,\,x_2,\,x_3) & \sim & \mathcal{N}(0.3 x_1 + 0.3 x_2 + 0.3 x_3,\,2^2).
\end{eqnarray*}
Further, we simulated a categorical variable $X_5$ with four categories using
\begin{eqnarray*}
P(X_5 = j\cond x_1, x_2, x_3, x_4) & = & \frac{\exp\{l_j\}}{\sum_{k=1}^4 \exp\{l_k\}},\
j=1,\ldots,4\\ \\
l_1 & = & 0,  \\
l_2 & = & -2 + 0.5 x_1 + 0.2 x_2 + 0.1 x_3 + 0.2 x_4, \\
l_3 & = & \hphantom{-} 0 + 0.5 x_1 + 0.2 x_2 + 0.1 x_3 + 0.2 x_4, \\
l_4 & = & -2 + 0.5 x_1 + 0.2 x_2 + 0.1 x_3 + 0.2 x_4, \\
\end{eqnarray*}
where the linear predictors $l_j$'s were chosen so that the first and the third categories were common (33\% and 53\%, respectively) while the second and the fourth categories had low frequencies (7\% each).

\subsubsection{Outcomes}
\label{sec: sim outcome}

The binary outcome $Y$ was simulated conditionally on the covariates using the following logistic distribution.
\begin{equation}
P(Y=1\cond x_1,\,x_2,\,x_3,\,x_4,\,x_5) = \frac{\exp(l)}{1+\exp(l)}, 
\end{equation}
where
\begin{multline}
l = -1 + 0.05\;x_1 + 0.2\;x_2 + 0.1\;x_3 + 0.02\;x_4 + \\ \log(5)\mathbf{1}(x_5=2) + \log(2)\mathbf{1}(x_5=3) + \log(1.5)\mathbf{1}(x_5=4),
\end{multline}
where $I(\cdot)$ denotes the indicator function with value 1 for true and 0 for false argument. For the categorical covariate $x_5$, two low frequent categories, category 4 had a low odds ratio (OR=1.5) while the other, the second had high (OR=5). These selections illustrate the bias related to low frequency categories, if they have a weak or strong association with the outcome. The third category, which had higher frequency, had OR=2.
The time-to-event outcome (TTE) $T^*$ was generated using the truncated Weibull model with the shape parameter $a$ and the scale parameter $b\exp\{-l_p/a\}$, and age as the time scale:
\begin{equation}
T^*\sim T \cond T>x_t, \; T \sim \text{Weibull}(a,\,b\exp\{-l_p/a\}).
\end{equation}

The survival function of $\text{Weibull}(\kappa, \lambda)$ is $S(t; \kappa, \lambda)  = exp(- (t/\lambda)^\kappa), t > 0$.

The distribution of baseline ages of the individuals is assumed to be $x_t\sim\text{Uniform}(0,50)$, and hence, we assumed delayed entries for our time-to-event analysis. This is a typical situation in population surveys, where the baseline age differs between individuals. 

In order to mimic a real life scenario, we used the population mortality statistics provided by Statistics Finland \citep{statfin_mortality}, to determine shape and scale parameters of the Weibull distribution. We used the estimates $\hat{a}=7.5$ and $\hat{b}=84$, respectively. 
The linear predictor was taken as
\begin{multline}
\label{eq:lp}
l_p =0.05\;x_1 + 0.2\;x_2 + 0.1\;x_3 + 0.02\;x_4 \\
  + \log(5)\;(x_5=2) + \log(2)\;(x_5=3) + \log(1.5)\;(x_5=4).
\end{multline}
The coefficients of the covariates $x_j$ can be interpreted as the log hazard ratios.
In this simulation study, we right censored the individuals at age 100 years. So, the observed TTE outcome variable $t_i$ and the event indicator $\delta_i$ for the $i^{\text{th}}$ individual or observation is based on the realized event time $t^*_i$, and the right censoring time 100, 
\begin{equation}
t_i = \min(t^*_i,\,100) \\
\end{equation}
\begin{equation}
\delta_i = \left\{
  \begin{array}{ll} 
      0 & \text{if the event was censored } (t^*_i > 100) \\
      1 & \text{if the event was observed } (t^*_i \leq 100). 
    \end{array}
\right.
\end{equation}
For the analyses in this section, we assumed that the variables  $x_1,\,x_2$ and the outcomes were fully observed for $\mathcal{P}\bigcup\mathcal{R}$. To assess the influence of missing data on the results, we considered four scenarios of $\mathcal{R}$: the values in $x_3$, $x_4$ and $x_5$ were only generated for 1\%, 5\%, 10\% or 20\% of $\mathcal{P}\bigcup\mathcal{R}$ assuming missing completely at random (MCAR). We generated 240 such data sets for the analyses.

\subsection{Evaluation of the missing data handling methods}

The imputed values are evaluated by convergence of iterations to ensure that the imputed values stabilize across iterations, which can be visualized through trace plots. Proper convergence is critical as it guarantees that the imputed values are consistent and reliable, leading to unbiased and valid statistical inferences in the presence of missing data. The MI method and CCA are compared based on the accuracy of the coefficient estimates $\beta^{k}_{\text{estimate}}$ for the $k^{\text{th}}$ of the $K$ simulated datasets with the true parameter $\beta_{\text{true}}$, using the following statistics:
\begin{description}
\item[Proportion of times when the estimate based on MI outperforms CCA] is defined as the average number of times when the estimate based on the MI method is closer to the true parameter than the estimate based on the CCA method,
\
    \begin{equation}
    \label{eq:r}
    d = \frac{1}{K} \sum_{k=1}^K \mathbf{1}\left\{|\hat\beta^{\text{MI}}_k - \hat\beta_{true} | < | \hat\beta^{\text{CCA}}_k - \hat\beta_{true} |\right\}, \\
    \end{equation}
    where $\hat\beta_k$ indicates the true value of the estimate when the data is complete without any missing values. $\hat\beta^{\text{MI}}_k$ denotes the estimates obtained by handling the missing data using MI technique and the $\hat\beta^{\text{CCA}}_k$ denotes the estimate obtained from CCA. 
\item[Mean absolute error (MAE)] is a measure of errors between paired observations expressing the same phenomenon, defined by
    \begin{equation}
    \label{MAE}
    \text{MAE} = \frac{1}{K}\sum_{k=1}^K\left| \beta_{\text{true}} - \beta^{k}_{\text{estimate}}\right|.
    \end{equation}
\item[Root mean square (RMSE)] is a quadratic scoring rule which also measures the mean magnitude of the error in the obtained parameter estimate, i.e. it’s the square root of the average of squared differences between estimated and actual parameter value.
    \begin{equation}
    \label{RMSE}
    \text{RMSE}=\sqrt{\frac{1}{K}\sum_{k=1}^K \left(\beta_{\text{true}} - \beta^{k}_{\text{estimate}}\right)^2}.
    \end{equation}

\end{description}

\section{Results}
\subsection{Simulation Study}
 
\subsubsection{Missing covariates in binary outcome data}
The simulation study reveals that MI outperforms CCA for binary outcomes, particularly when dealing with partially observed covariates $x_3$, $x_4$ and $x_5$. The effectiveness of MI compared to CCA is described through the closer alignment of MI estimates with true values across all scenarios, demonstrated in Table~\ref{tab:ratio1}. 
 
In this case of the logistic regression model, that in more than 50\% times MI performs better than CCA in all cases after handling missing values in $x_3$, $x_4$, and $x_5$ (Table~\ref{tab:ratio1}). Nonetheless, MI's performance for the variable $x_3$ and the second category of the categorical variable $x_5$ is slightly weaker, likely due to the low occurrence rate of this category within $x_5$.

\begin{table}[H]
\caption{Proportion of times when the estimate based on MI outperforms CCA ($d$, equation~\eqref{eq:r}) by the percentages of observed data.}
\label{tab:ratio1}
\centering
\renewcommand{\arraystretch}{1.5}
\begin{tabular}{lrrrrr}
\toprule
Observed (\%) in $\mathcal{P}\bigcup\mathcal{R}$ & $x_3$ & $x_4$ & $x_5^{(2)}$ & $x_5^{(3)}$ & $x_5^{(4)}$ \\
\midrule
1\%  & 0.83 & 0.69 & 0.83 & 0.83 & 0.55 \\
5\%  & 0.63 & 0.79 & 0.63 & 0.87 & 0.65 \\
10\% & 0.53 & 0.79 & 0.53 & 0.86 & 0.69 \\
20\% & 0.59 & 0.77 & 0.59 & 0.77 & 0.77 \\
\bottomrule
\end{tabular}
\end{table}

MI's superiority over CCA becomes more evident with an increase in missing data, as shown by the RMSE and the MAE analyses in Figures~\ref{fig:RMSE} and~\ref{fig:MAE}, respectively. Even for the second category of $x_5$, MI outperforms CCA at extremely high missingness levels (99\% and 95\%).

\begin{figure}[H]
\centering
\begin{subfigure}[H]{.95\linewidth}
\includegraphics[width=\linewidth]{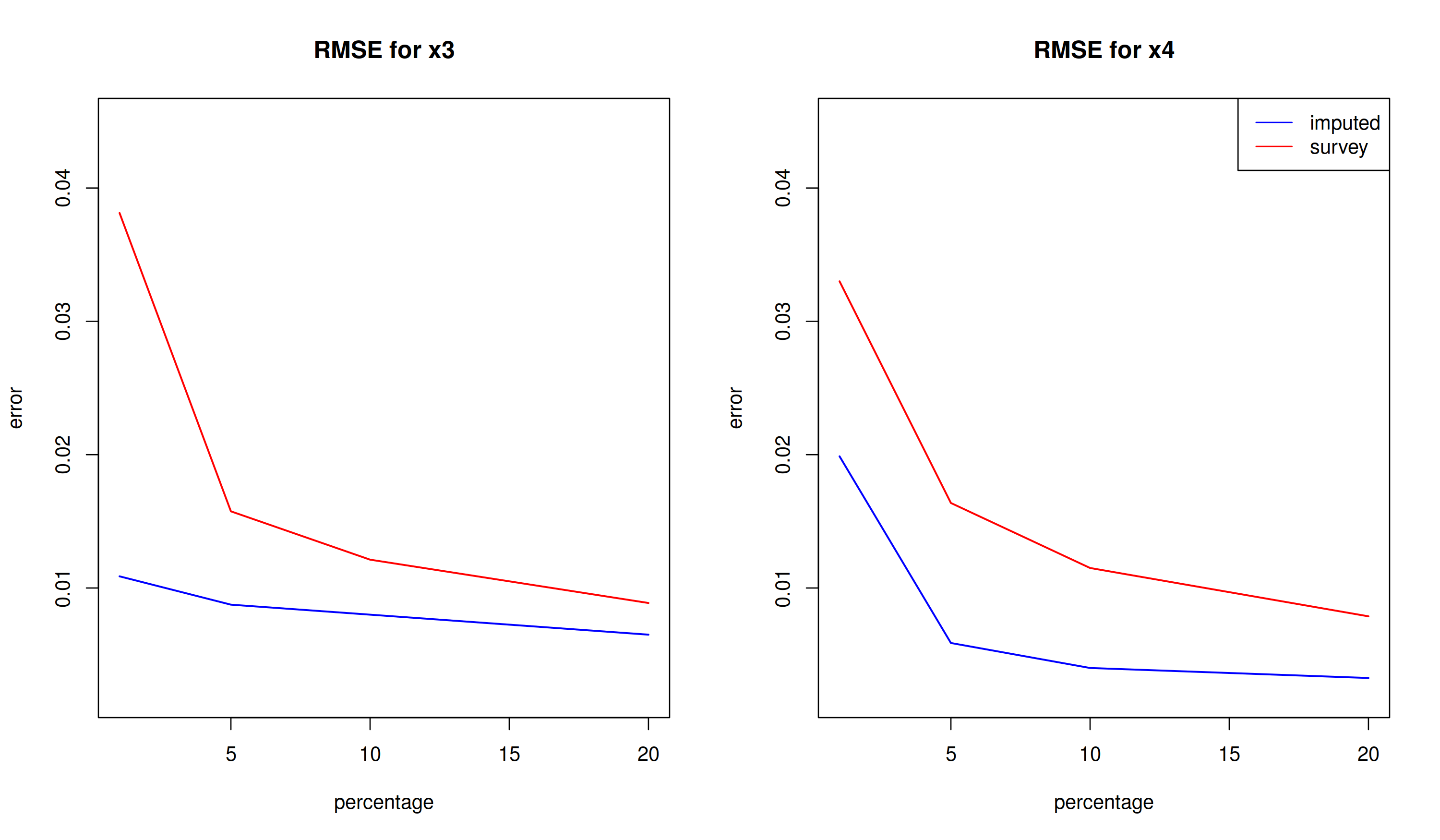}
\caption{RMSE of regression coefficients of $x_3$ (left panel) and $x_4$ (right panel) for a binary outcome, with comparisons between MI (blue) and CCA (red) methods.}
\label{fig:RMSE(a)}
\end{subfigure}

\begin{subfigure}[H]{.95\linewidth}
\includegraphics[width=\linewidth]{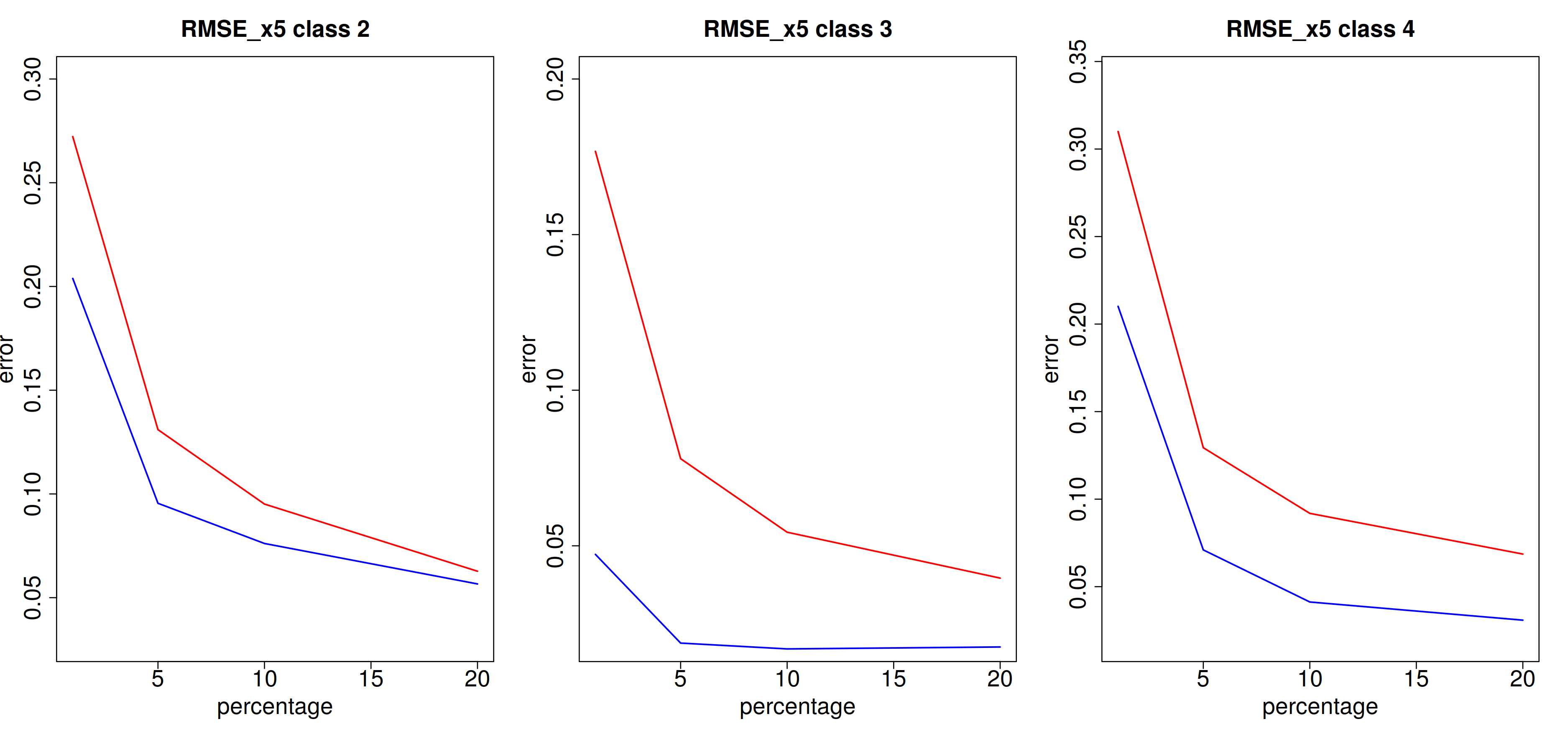}
\caption{RMSE of regression coefficients for the three categories of $x_5$, with comparisons between MI (blue) and CCA (red) methods.}
\label{fig:RMSE(b)}
\end{subfigure}
\caption{Root mean square error (RMSE, equation~\eqref{RMSE}) of the regression coefficients in the case of a binary outcome.}
\label{fig:RMSE}
\end{figure}

\begin{figure}[H]
\centering
\begin{subfigure}[H]{.95\linewidth}
\includegraphics[width=\linewidth]{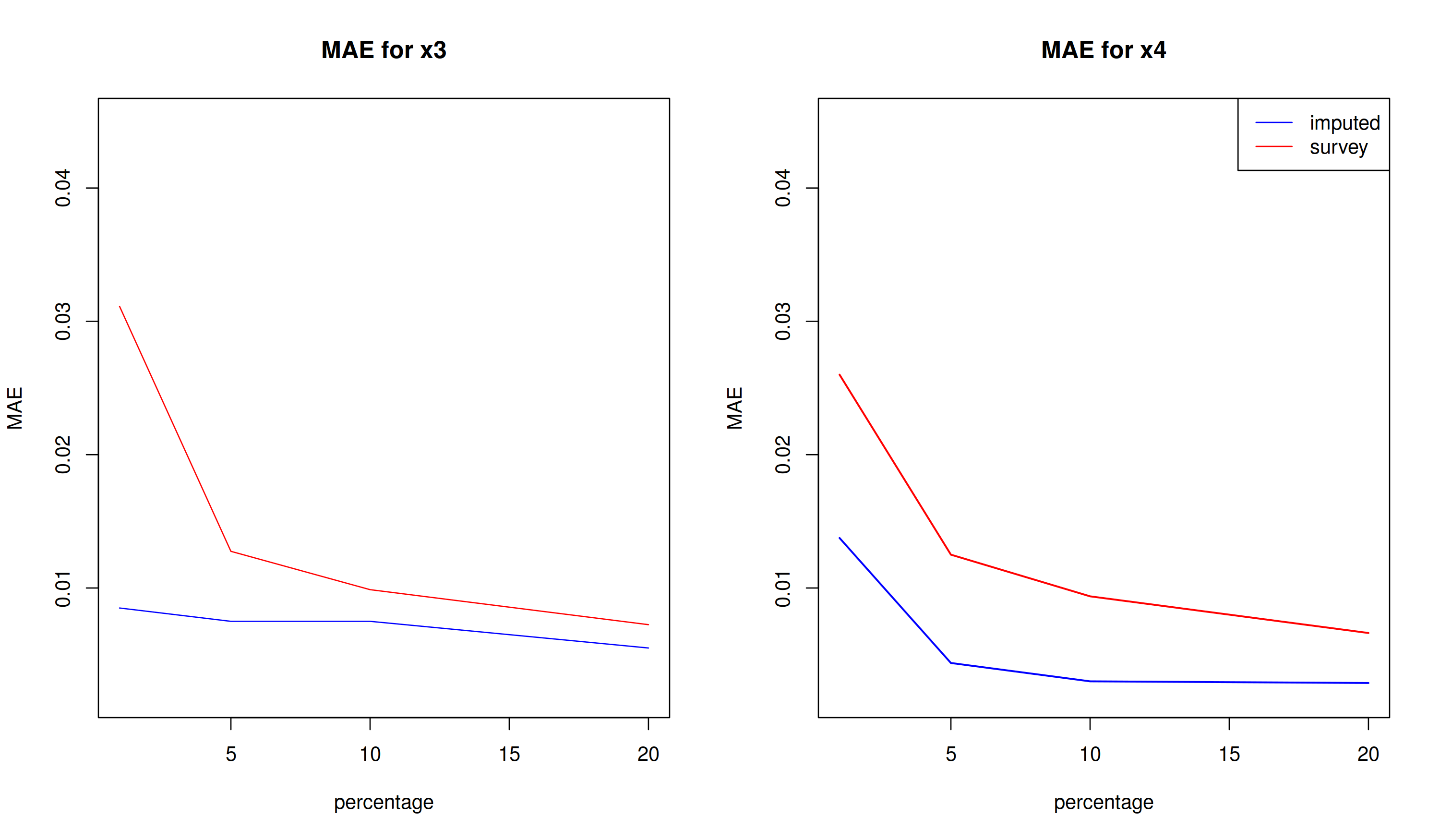}
\caption{MAE of regression coefficients of $x_3$ (left panel) and $x_4$ (right panel) for a binary outcome, with comparisons between MI (blue) and CCA (red) methods.}
\end{subfigure}

\begin{subfigure}[H]{.95\linewidth}
\includegraphics[width=\linewidth]{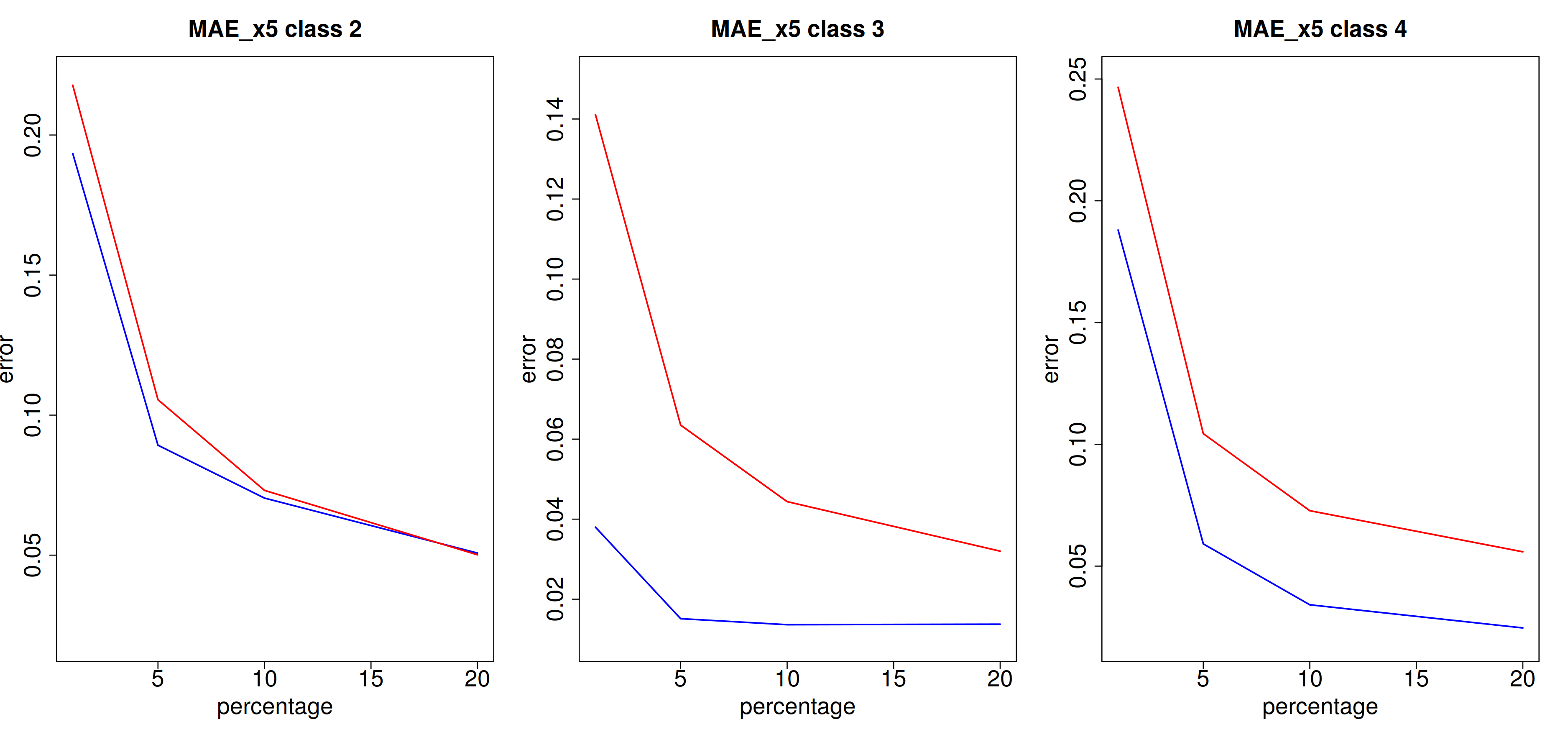}
\caption{MAE of regression coefficients for the three categories of $x_5$, with comparisons between MI (blue) and CCA (red) methods.}
\end{subfigure}
\caption{Mean absolute errors (MAE, equation~\eqref{MAE}) of the regression coefficients in case of a binary outcome for different percentage of missing data.}
\label{fig:MAE}
\end{figure}

\subsubsection{Missing covariates in the time-to-event outcome}

The results demonstrate that for the time-to-event survival outcome variable~$y = (T,\delta)$, generated alongside covariates $x_1$ and $x_2$ (without missing values) and $x_3$, $x_4$, and $x_5$ (with observed values ranging from 1\% to 20\%), handled using MI is comparable with the CCA.
Table~\ref{tab:ratio2} demonstrate that the parameter $d$ denoting the proportion of times estimates obtained from the MI data outperform CCA are below 0.5, suggesting that CCA estimates are often closer to the actual values compared to those obtained by MI.

\begin{table}[H]
\caption{Comparison of MI and CCA methods in terms of their proximity $d$ to the true value, calculated as per equation \eqref{eq:r}.}
\renewcommand{\arraystretch}{1.5}
\begin{tabular}{lcccccc}
\hline
MI Method & Observed(\%) in $\mathcal{P}\bigcup\mathcal{R}$ & $x_3$ & $x_4$ & $x_5^{(2)}$ & $x_5^{(3)}$ & $x_5^{(4)}$ \\
\hline
CART & 1\% & 0.35 & 0.48 & 0.1 & 0.49 & 0.48 \\
     & 5\% & 0.32 & 0.35 & 0.08 & 0.46 & 0.43 \\
     & 10\% & 0.32 & 0.38 & 0.05 & 0.47 & 0.41 \\
     & 20\% & 0.28 & 0.39 & 0.03 & 0.40 & 0.35 \\
GLM  & 1\% & 0.34 & 0.41 & 0.38 & 0.54 & 0.52 \\
     & 5\% & 0.13 & 0.30 & 0.32 & 0.43 & 0.47 \\
     & 10\% & 0.12 & 0.36 & 0.25 & 0.40 & 0.48 \\
     & 20\% & 0.08 & 0.37 & 0.16 & 0.35 & 0.46 \\
Transformation & 1\% & 0.38 & 0.47 & 0.11 & 0.51 & 0.49 \\
               & 5\% & 0.32 & 0.39 & 0.06 & 0.48 & 0.39 \\
               & 10\% & 0.32 & 0.38 & 0.06 & 0.46 & 0.40 \\
               & 20\% & 0.24 & 0.40 & 0.03 & 0.46 & 0.36 \\
\hline
\end{tabular}
\label{tab:ratio2}
\end{table}

Despite this, analyses of Root Mean Square Error (RMSE) and Mean Absolute Error (MAE) indicate that the MI and CCA methods perform comparably in estimating the coefficients, as depicted in Figures~\ref{fig: Survival RMSE} and \ref{fig:Survival MAE}.

\begin{figure}[H]
\centering
\begin{subfigure}[H]{.95\linewidth}
\includegraphics[width=\linewidth]{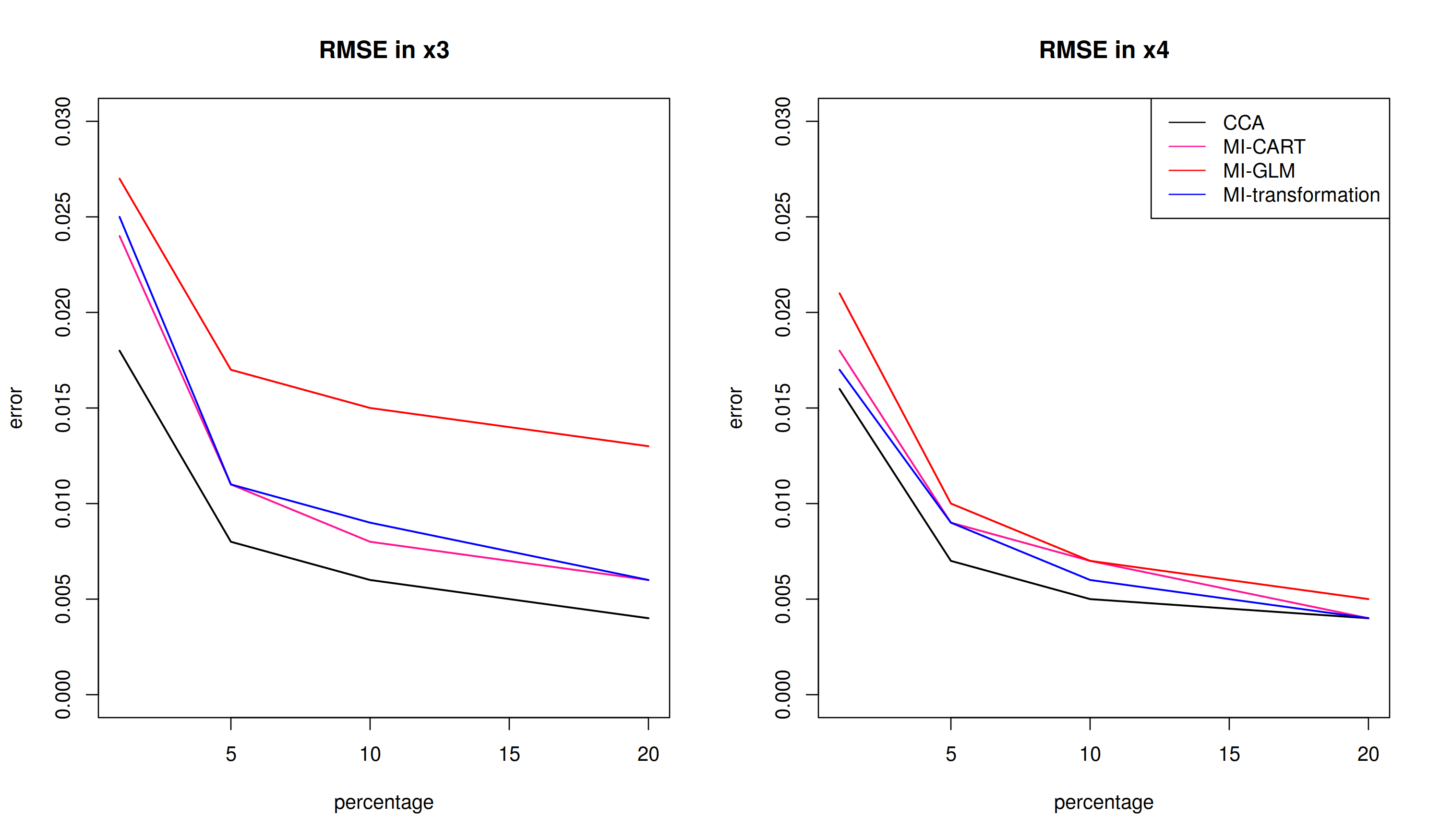}
\caption{RMSE of regression coefficients of $x_3$ (left panel) and $x_4$ (right panel) for a survival outcome with comparisons between MI with imputation model CART (deep pink), GLM (red), transformation (blue) and CCA (black) methods
}
\label{fig: Survival RMSE(a)}
\end{subfigure}

\begin{subfigure}[H]{.95\linewidth}\includegraphics[width=\linewidth]{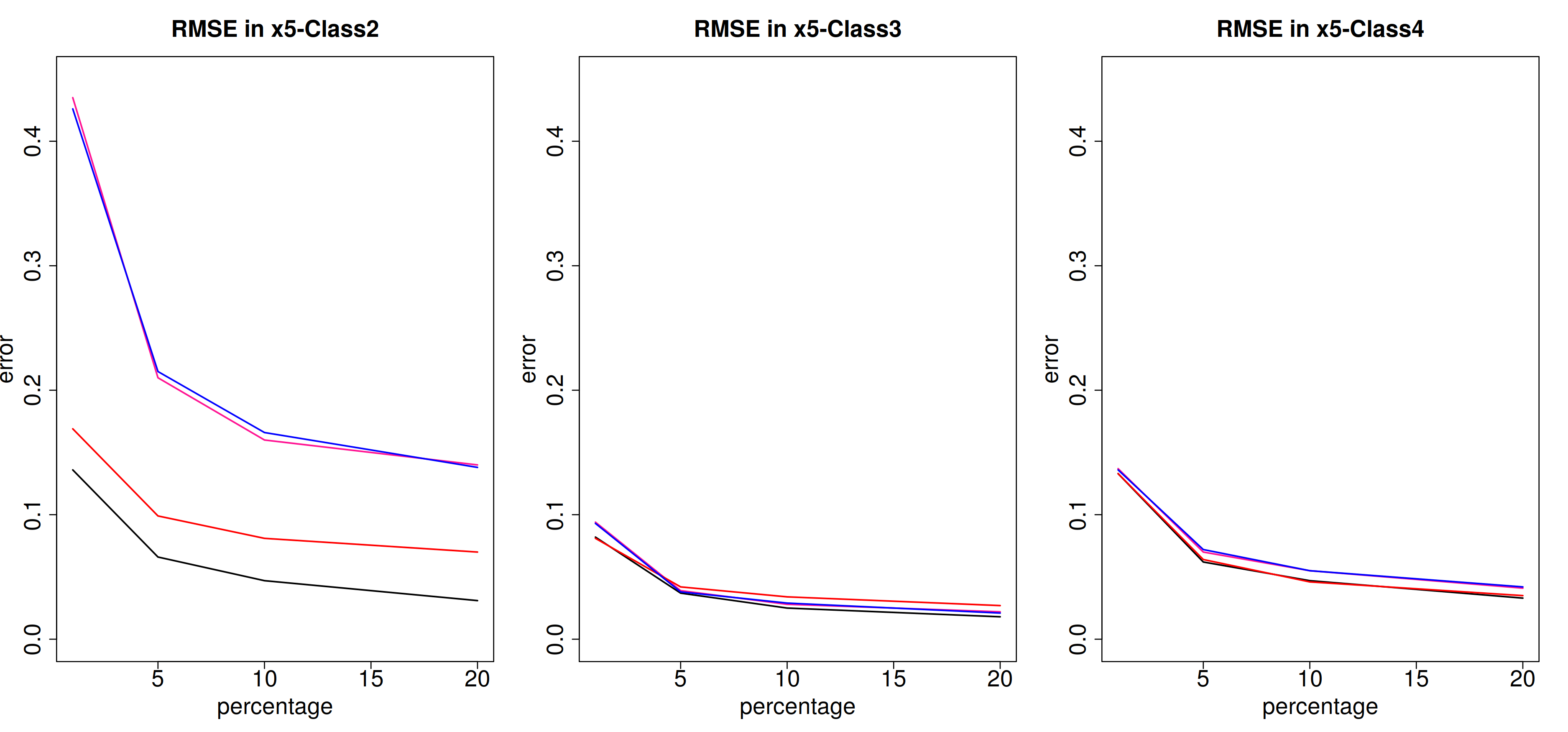}
\caption{RMSE of regression coefficients for the three categories of $x_5$, with comparisons between MI with imputation model CART (deep pink), GLM (red), transformation (blue) and CCA (black) methods.
}
\label{fig: Survival RMSE(b)}
\end{subfigure}
\caption{Root mean square error (RMSE, equation~\eqref{RMSE}) of the regression coefficients in case of a binary outcome for different percentage of missing data.
}
\label{fig: Survival RMSE}
\end{figure}

\begin{figure}[H]
\centering
\begin{subfigure}[H]{.95\linewidth}
\includegraphics[width=\linewidth]{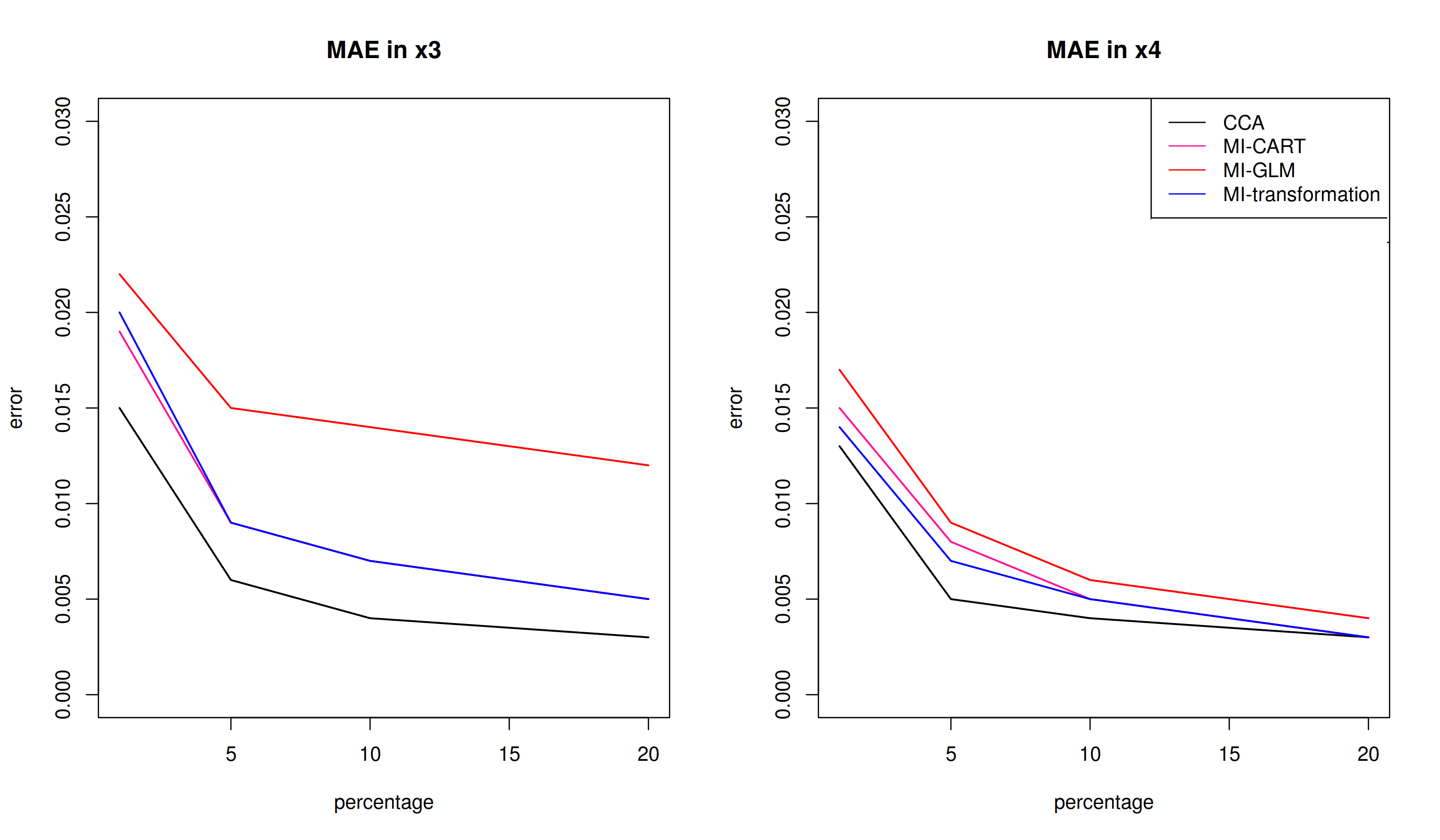}
\caption{MAE of regression coefficients of $x_3$ (left panel) and $x_4$ (right panel) for a survival outcome with comparisons between MI with imputation model CART (deep pink), GLM (red), transformation (blue) and CCA (black) methods.
}
\label{fig: Survival MAE(a)}
\end{subfigure}

\begin{subfigure}[H]{.95\linewidth}\includegraphics[width=\linewidth]{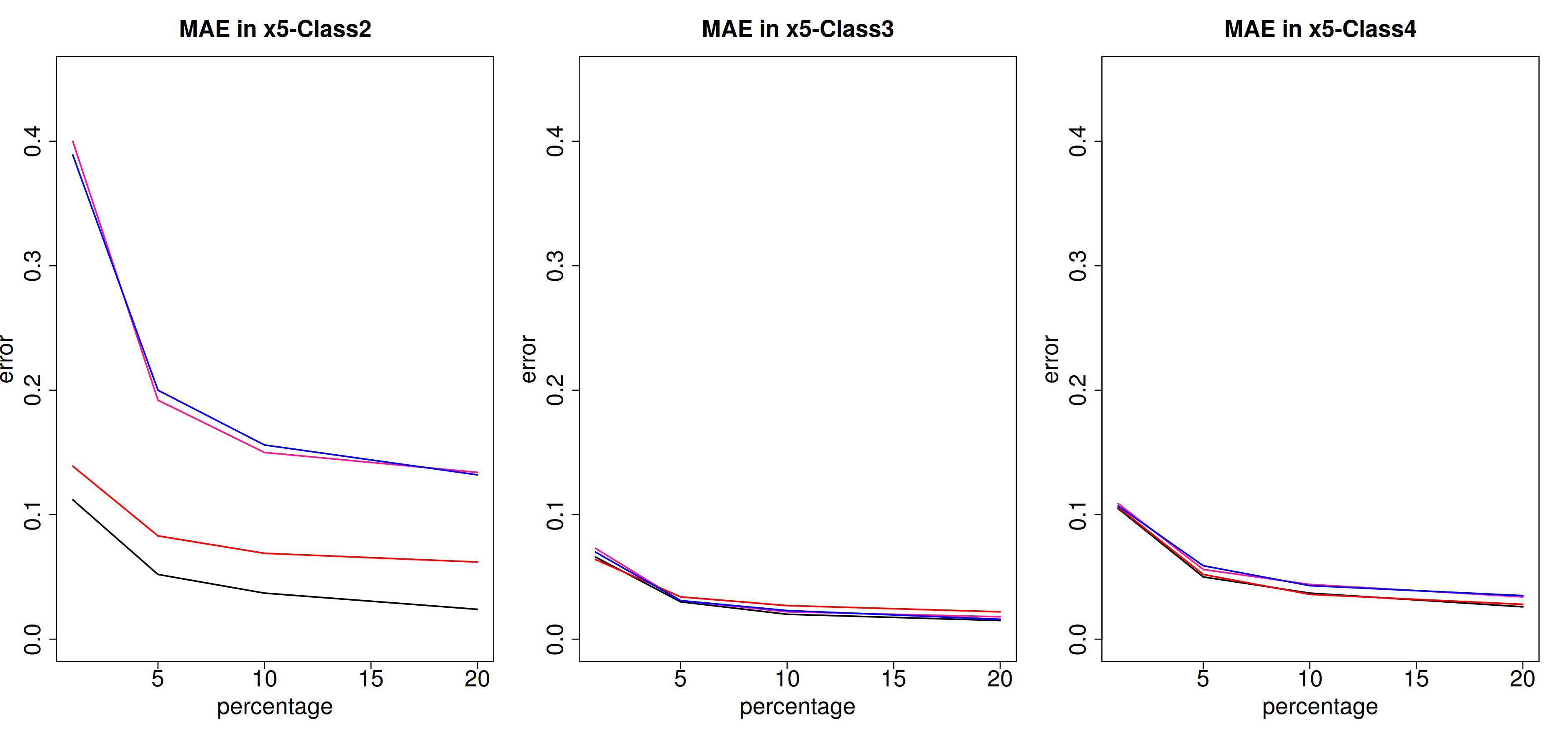}
\caption{MAE of regression coefficients for the three categories of $x_5$, with comparisons between MI with imputation model CART (deep pink), GLM (red), transformation (blue) and CCA (black) methods.
}
\label{fig: Survival MAE(b)}
\end{subfigure}
\caption{Mean absolute errors (MAE, equation~\eqref{MAE}) of the regression coefficients in case of a binary outcome for different percentage of missing data.
}
\label{fig:Survival MAE}
\end{figure}

\subsection{Empirical study}

The complete case data based on covariates with no missing values contains 4,792 subjects which is 65\% of the survey sample (Table~\ref{tab:event}). The population data are more than 100 times larger ($N = 492,421$), but the numbers of IHD cases and non-IHD deaths are 150-fold. This indicates that IHD is more prevalent among nonparticipants. In the survey sample the proportions are similar to the population data as expected, because the selective nonparticipation does apply both in the survey sample and in the population data.

\begin{table}
\caption{The descriptive statistics of the empirical data. (N = number of individuals).}
\label{tab:event}
\centering
\renewcommand{\arraystretch}{1.5}
\begin{tabular}{@{}lrrrr@{}}
\toprule
Data & N & IHD cases & \begin{tabular}[c]{@{}l@{}}Non IHD\\ death\end{tabular} & Censored (\%) \\ \midrule
CCA $\mathcal{R}$ & 4,792 & 575 & 456 & 3,761 (78\%)\\
Survey $\mathcal{C}$ & 7,938 & 1,282 & 1,306 & 5,349 (67\%) \\
Population and survey $\mathcal{P}\cup\mathcal{C}$ & 492,421 & 91,073 & 71,651 & 329,697 (67\%) \\ \bottomrule
\end{tabular}
\end{table}

\begin{figure}
\begin{subfigure}[h]{.49\linewidth}
\includegraphics[width=\linewidth,height=3.25in]{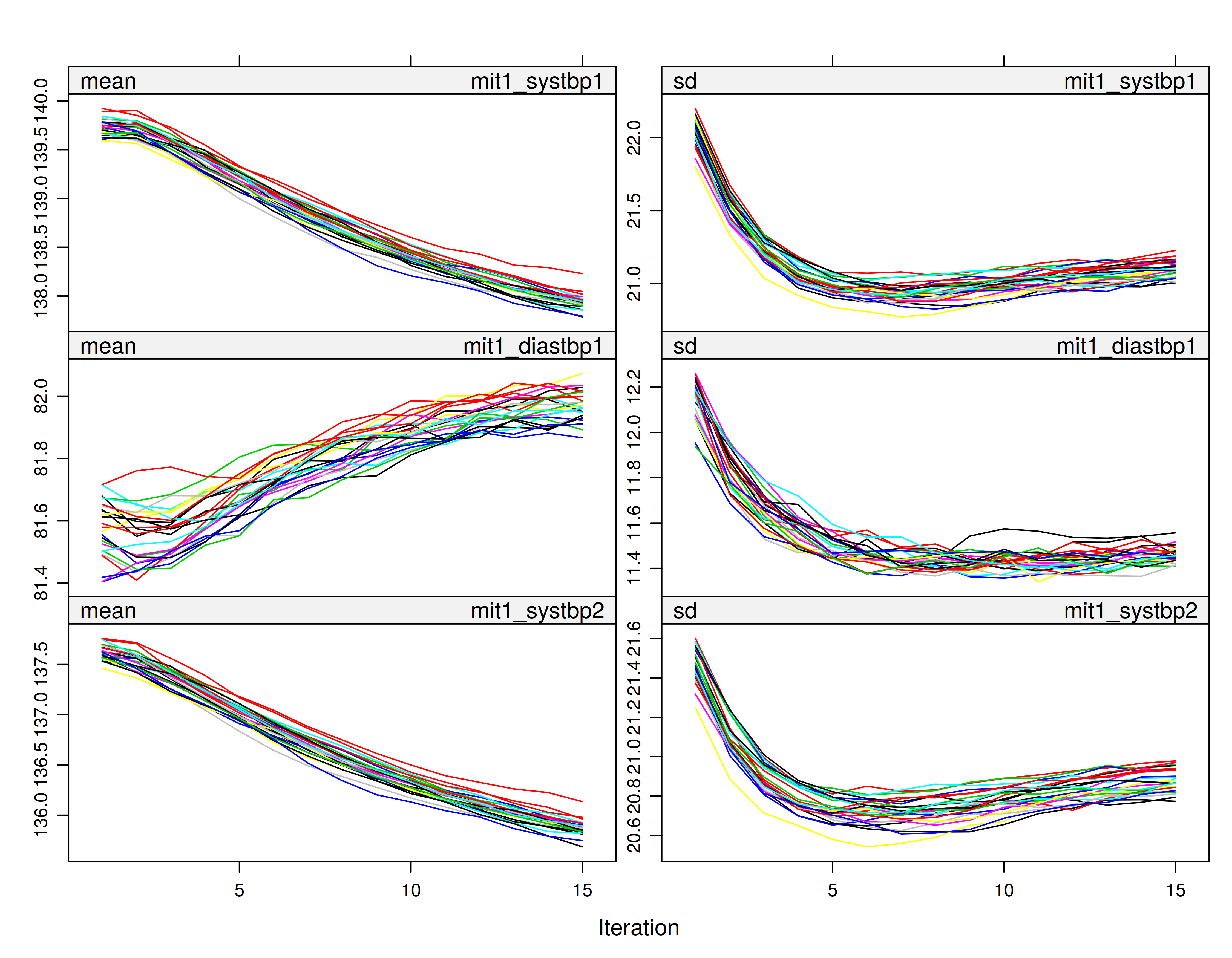}
\caption{1st measurement of SBP and DBP, \newline 2nd measurement of SBP}
\label{fig:conv(a)}
\end{subfigure}
\begin{subfigure}[H]{.49\linewidth}
\includegraphics[width=\linewidth,height=3.25in]{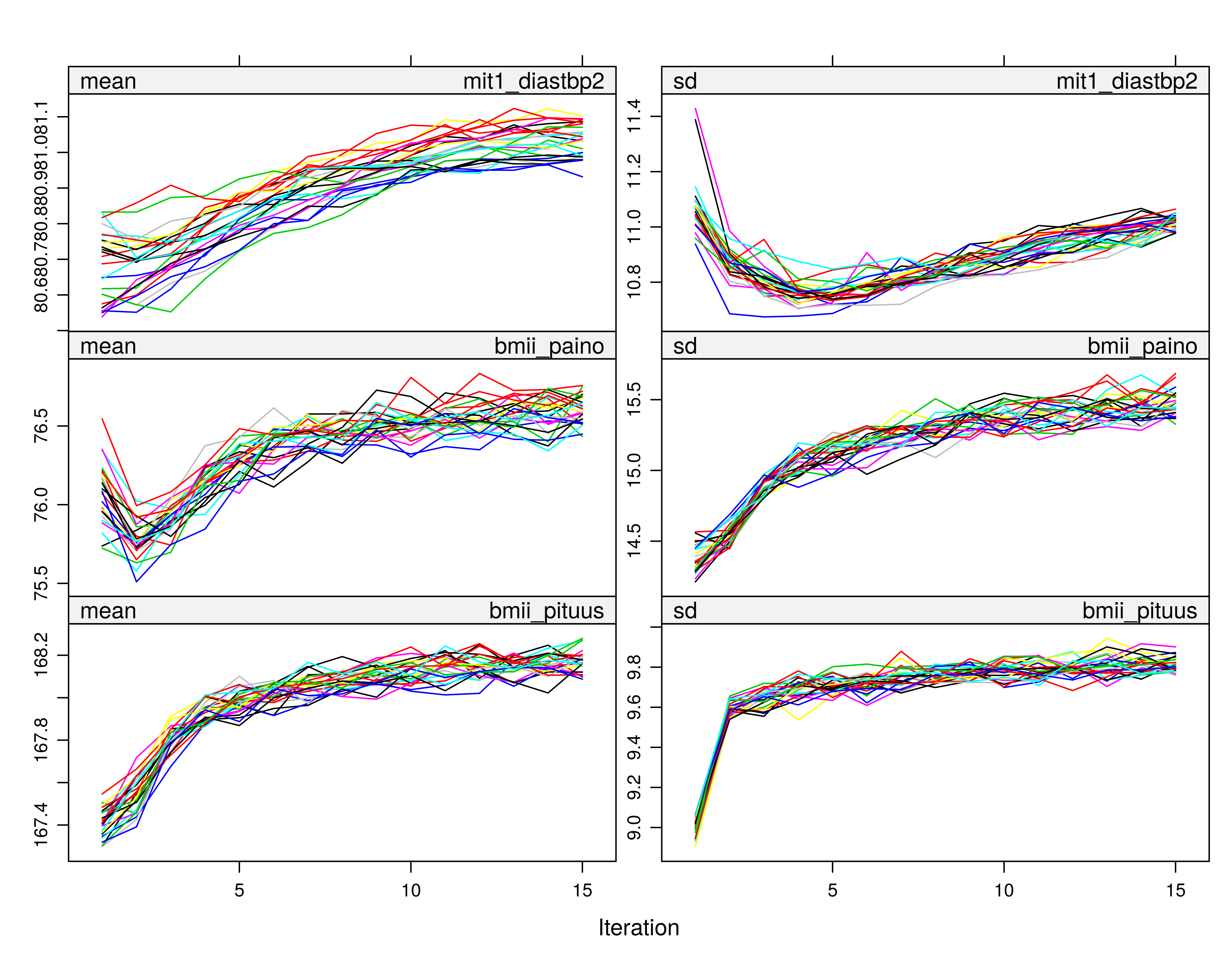}
\caption{2nd DBP measurement, weight (kg), height (cm)}
\label{fig:conv(b)}
\end{subfigure}
\centering
\begin{subfigure}[H]{.59\linewidth}
\includegraphics[width=\linewidth,height=3.25in]{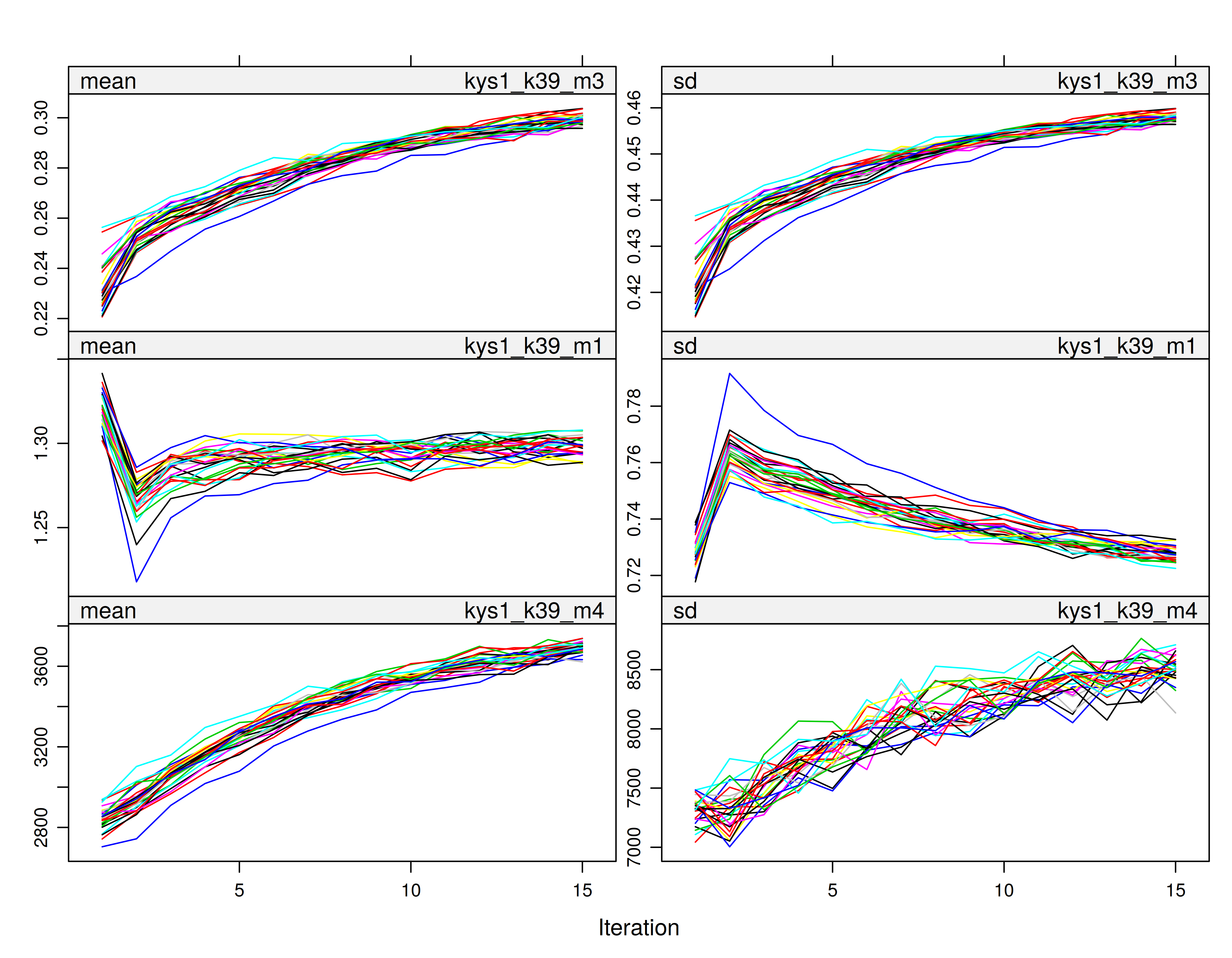}
\caption{Alcohol consumption; high-risk indicator, consumption frequency and quantity (g/year)} 
\label{fig:conv(c)}
\end{subfigure}
\caption{Trace plots of nine imputed covariate values obtained using MICE method for the combined samples. The left- and right-hand side plots in each panel show the mean and the standard deviation of imputed values of the missing variables, respectively. The curves show the changes over the 15 iterations of the MICE algorithm. The variables in each panel are given from top to bottom. (SBP: systolic blood pressure, DBP: diastolic blood pressure)}
\label{fig:Conv}
\end{figure}

After imputing the missing values using the MICE method in the data combining survey sample with population data, convergence of each of the imputed covariates was assessed. The trace plots of nine imputed covariates in the population data, in general, show slow or no convergence (Figure~\ref{fig:Conv}). 
The standard deviations (Figure~\ref{fig:Conv}, right-hand side plots of each panel) of the imputed covariate values seem to stabilize faster than their means (Figure~\ref{fig:Conv}, left-hand side plots of each panel).
\begin{table}
\centering
\renewcommand{\arraystretch}{1.5}
\Rotatebox{90}{%
\scalebox{0.8}{
\begin{tabular}{llllllllllllll} 
\cline{1-2}\cline{3-14}
\multirow{4}{*}{}                                                                   & \multirow{4}{*}{} & \multicolumn{6}{l}{Transition from state 1 to 2}                                                                                  & \multicolumn{6}{l}{Transition from state 1 to 3}                                                                                   \\ 
\cline{3-14}
                                                                                    &                   & \multicolumn{3}{l}{Estimates}                                   & \multicolumn{3}{l}{Std error}                                   & \multicolumn{3}{l}{Estimates}                                   & \multicolumn{3}{l}{Std error}                                    \\ 
\cline{3-14}
                                                                                    &                   & \multicolumn{2}{c}{MI}                         & CCA            & \multicolumn{2}{c}{MI}                         & CCA            & \multicolumn{2}{c}{MI}                         & CCA            & \multicolumn{2}{c}{MI}                         & CCA             \\
                                                                                    &                   & $\mathcal{P}\cup\mathcal{C}$  & $\mathcal{C}$  & $\mathcal{R}$  & $\mathcal{P}\cup\mathcal{C}$  & $\mathcal{C}$  & $\mathcal{R}$  & $\mathcal{P}\cup\mathcal{C}$  & $\mathcal{C}$  & $\mathcal{R}$  & $\mathcal{P}\cup\mathcal{C}$  & $\mathcal{C}$  & $\mathcal{R}$   \\ 
\cline{1-2}\cline{3-14}
Gender                                                                              & 2                 & -0.069                        & -0.678         & -1.069         & 0.010                         & 0.042          & 0.135          & -0.032                        & 0.157          & -0.705         & 0.012                         & 0.087          & 0.151           \\
Regular smoking                                                              & 2                 & 0.211                         & 0.152          & 0.095          & 0.008                         & 0.039          & 0.116          & 0.249                         & 0.644          & 0.601          & 0.010                         & 0.076          & 0.117           \\
Total cholesterol                                                                   &                   & -0.064                        & -0.029         & -0.094         & 0.003                         & 0.013          & 0.04           & -0.086                        & -0.11          & -0.054         & 0.004                         & 0.027          & 0.044           \\
HDL cholesterol                                                                     &                   & -0.505                        & -0.603         & -0.617         & 0.010                         & 0.044          & 0.14           & -0.326                        & -0.135         & -0.072         & 0.011                         & 0.08           & 0.145           \\
Systolic BP1                                                                        &                   & 0.001                         & -0.001         & -0.004         & 0.001                         & 0.002          & 0.007          & 0.002                         & -0.002         & -0.010         & 0.001                         & 0.004          & 0.008           \\
Diastolic BP1                                                                       &                   & 0.002                         & -0.001         & -0.019         & 0.001                         & 0.002          & 0.011          & -0.001                        & 0.006          & 0.018          & 0.001                         & 0.005          & 0.013           \\
Systolic BP2                                                                        &                   & -0.004                        & 0.004          & 0.014          & 0.001                         & 0.002          & 0.007          & -0.005                        & -0.003         & 0.018          & 0.001                         & 0.004          & 0.008           \\
Diastolic BP2                                                                       &                   & 0.009                         & -0.008         & 0.006          & 0.001                         & 0.003          & 0.011          & 0.004                         & 0.009          & -0.026         & 0.001                         & 0.006          & 0.013           \\
Weight                                                                              &                   & -0.005                        & 0.004          & 0.010          & 0.000                         & 0.001          & 0.004          & -0.006                        & -0.019         & -0.003         & 0.000                         & 0.003          & 0.004           \\
Height                                                                              &                   & 0.013                         & -0.012         & -0.039         & 0.001                         & 0.002          & 0.008          & 0.017                         & 0.012          & -0.02          & 0.001                         & 0.005          & 0.009           \\
Physical activity                                                        & 2                 & -0.019                        & -0.144         & -0.181         & 0.007                         & 0.03           & 0.096          & -0.413                        & -0.365         & -0.409         & 0.008                         & 0.061          & 0.105           \\
                                                                                    & 3                 & -0.068                        & -0.319         & -0.241         & 0.011                         & 0.048          & 0.138          & -0.304                        & -0.367         & -0.35          & 0.012                         & 0.098          & 0.154           \\
                                                                                    & 4                 & -0.093                        & -0.511         & -0.468         & 0.034                         & 0.176          & 0.456          & -0.44                         & -0.715         & -0.946         & 0.043                         & 0.413          & 0.585           \\
Alcohol usage                                                                       & 2                 & 0.028                         & 0.122          & -0.342         & 0.016                         & 0.054          & 0.394          & 0.075                         & 0.352          & 0.224          & 0.019                         & 0.106          & 0.311           \\
                                                                                    & 3                 & 0.002                         & -0.147         & -0.077         & 0.014                         & 0.057          & 0.212          & 0.032                         & -0.327         & -0.819         & 0.016                         & 0.112          & 0.202           \\
Past year alcohol use                                                   &                   & 0.031                         & -0.003         & -0.026         & 0.001                         & 0.004          & 0.012          & 0.041                         & 0.057          & 0.013          & 0.001                         & 0.007          & 0.012           \\
Over use of alcohol                                                                 & 2                 & 0.039                         & -0.036         & -0.139         & 0.009                         & 0.05           & 0.125          & -0.022                        & 0.127          & 0.037          & 0.011                         & 0.100          & 0.132           \\
\begin{tabular}[c]{@{}l@{}}Past year frequency \\ of alcohol use\end{tabular} & 2                 & -0.168                        & -0.208         & -              & 0.016                         & 0.068          &                & 0.077                         & 0.121          & -              & 0.018                         & 0.133          &                 \\
                                                                                    & 3                 & -0.009                        & 0.050          & 0.175          & 0.014                         & 0.055          & 0.108          & 0.099                         & 0.073          & 0.066          & 0.016                         & 0.107          & 0.122           \\
Total alcohol g/year                                              &                   & 0.000                         & 0.004          & 0.003          & 0.000                         & 0.001          & 0.002          & 0.000                         & 0.001          & 0.005          & 0.000                         & 0.001          & 0.001           \\
\cline{1-1}\cline{3-14}
\end{tabular}}}
\caption{Ischemic heart disease data: results for the parameter estimate in the analysis model, where the imputation model is chosen based on the Kendall rank correlation method. $\mathcal{P}\cup\mathcal{C}$ corresponds to the combined population and survey data,  $\mathcal{C}$ to the survey data, and $\mathcal{R}$ to compete cases (CCA). BP is blood pressure.}
\label{tab:est}
\end{table}

The coefficient estimates obtained from MI on the survey cohort were closer to CCA than that obtained obtained from the MI using the population data (Table~\ref{tab:est}). The standard errors for the estimates were much lower for the MI method using the population data, when compared to the other two estimates.

\newpage
\section{Discussion}
In this paper, we have implemented three distinct analyses, utilizing complete cases $\mathcal{R}$  from the survey, all data from the survey $\mathcal{C}$, and all data from the survey and a random sample of the population $\mathcal{P} \cup \mathcal{C}$. 
In scenarios devoid of missing survey data, the first and the second options are the same. For the second and the third analytical approaches, MI techniques were utilized to address missing data problems. 
We have allowed possible non-response for the survey sample, which is important as the response rates have been declining over the past years. The augmented sample can be considered as a health survey sample with a large number of subjects and with missing risk factor data. The missingness mechanism in the sample was of mixed type; MNAR or MAR for $\mathcal{C} \setminus \mathcal{R}$ while MCAR for $\mathcal{P}$, as in the latter, missingness can be controlled by the researcher. Such mechanisms are justifiable considering the knowledge of the non-response mechanism and the selection of $\mathcal{P}$ without reference to the risk factors or other factors often associated with non-participation in population surveys. The proportions of cases (IHD $16\%, 18\%$, non-IHD death $16\%, 15\%$) and censoring ($67\%$, $67\%$) were similar between the original survey sample $\mathcal{C}$ and the augmented sample $\mathcal{P} \cup \mathcal{C}$. Among the survey participants $\mathcal{R}$, the proportions of cases were lower, $12\%$ (IHD) and $10\%$ (non-IHD death), and the proportion of censoring was higher, $78\%$. This may suggest that the survey participants are likely to be healthier, which is often the case in non-participation (\cite{Jousilahti2005}). It is to be noted that these are unadjusted results, so any differences in the age distributions between participants and non-participants can influence these results. 

Recently a supersampling approach for nested case-control and case-cohort studies, in which a nested case-control or case-cohort sample is supplemented with additional controls has been proposed and two MI methods have been studied \citep{borgan2023}. Our approach of augmenting the survey  sample can be seen as a supersampling approach though there is a subtle difference between these two ideas. A supersample increases  the number of controls per case in a nested case-control study or the size of the subcohort in a case-cohort study. Both these type of studies include all cases from the target population observed during the study period. In our sample augmentation approach, we only increase the size of the sample with the aim to gain efficiency in estimating the association parameters. The philosophy behind the case-cohort study is to augment the subcohort by all cases from the original cohort and collect some expensive covariates for the case-cohort set. In our setting, we could augment the survey sample by all cases from the target population. This would cause imbalance between the proportions of cases in the augmented sample and the survey sample, and analyses approaches such as the inverse-probability weighting are often used in such settings. This and other augmentation approaches and analyses methods would be of interest to investigate in future.

We carried out simulation studies to assess the imputation methods for analyses of augmented samples with varying degree of association between the risk factors and the outcome. We also considered continuous and categorical risk factors with rare categories.  From the simulation studies, it was evident that in the case of a binary outcome, the best method for imputing all types of variables considered is the MI, as standard imputation methods can handle these distributions with little difficulties. These observations apply to data with a small proportion of observed values (1\% - 20\%). However, in the case of time-to-event outcomes, it was seen that the accuracy of the coefficient estimates are almost the same for both CCA and MI for data with a small proportion of observed values (1\% - 20\%). A possible explanation is that the successful application of a large, additional population sample can be sensitive to the choice of the imputation model, when the large majority of the risk factor values are missing, and is in concordance with the results of \citep{borgan2023} for the large superset sample. 

The convergence problems (Figure~\ref{fig:Conv}), which we encountered, also illustrated that the initial values for the missing data were poor. This could be improved by using a one-chain approach, in which a long burn-in could help finding more realistic values for the missing data, and after that collecting every, say, fifth iterations as the imputed datasets to reduce the otherwise overwhelming computational burden.
Furthermore, many variables exhibit distributions that are not easily captured by standard parametric models. This limitation has led to the increased use of non-parametric methods such as CART for conditional imputation (\cite{assmann2014nonparametric} and \cite{doi:10.1080/00031305.2016.1277158}). CART's flexibility in handling complex, nonlinear relationships makes it a valuable tool for MI. \cite{doi:10.1080/00031305.2016.1277158} demonstrated that default chained equations approaches based on generalized linear models (GLMs) are often outperformed by default regression tree and Bayesian mixture models for the imputation of categorical data. Even though from our results there is no clear winner of the imputation model for all type of variables.

According to \cite{royston2009multiple}, when dealing with missing values in a binary or normally distributed variable $X$ in survival outcome data, the imputation model typically involves logistic or linear regression on the event indicator $\delta$, the cumulative baseline hazard (Nelson-Aalen estimate), and other remaining covariates for doing MI. This approach, however, can become problematic in datasets with a common outcome event and a large number of variables, especially when relationships among variables are interactive and nonlinear (\cite{Casiraghi2023}).  In the case of imputation of both continuous and categorical variables with MI, all the imputation models including the one suggested by \cite{white2009imputing}, CART, and imputation model containing survival time $T$ and its transformation $\log T$ along with the event indicator $\delta$ (\cite{vanBuuren1999}) with remaining variables had comparatively similar efficiency in terms of error values.

Our results clearly indicate the potential in using large, additional population samples, which are generally inexpensive to collect, to improve the accuracy of the estimates of interest for a binary outcome data. Our findings suggest that tree-based methods might not be optimal for imputing data with time-to-event variables, and parametric models could provide more transparent approach to impute the missing values. 
There is a pressing need for further research, especially for a large proportion of missing risk factor values in time-to-event designs.
 
In principle, we could analyse the entire target population under the mixed types of missing data mechanism. The gain in doing so may not be sufficient considering the computation time due to the missing risk factor data for almost the entire cohort. In association studies of rare events, supplementing a survey sample with a random sample of the target population is a viable option. 

\appendix 

\section{Appendix}

The \texttt{nelsonaalen} function from the package \texttt{mice} calculates the estimate of cumulative hazard values for each individuals. However, this function does not incorporate delayed entries. It assumes the starting time for all individual as 0. 

We have modified the \texttt{nelsonaalen} function that estimates the cumulative hazard in case of delayed entries. The following is the R code for the same.

\begin{verbatim}
##############################################################
                   Nelson-Aalen estimate
##############################################################                                              
nelson_aalen <- function(data, timevar, statusvar, starttime) {
  mice:::install.on.demand("survival")
  if (!is.data.frame(data)) 
    stop("Data must be a data frame")
  timevar <- as.character(substitute(timevar))
  statusvar <- as.character(substitute(statusvar))
  time <- data[, timevar]
  status <- data[, statusvar]
  hazard1 <- survival::
    basehaz(survival::coxph(survival::Surv(time, status) ~ 1))
  idx1 <- match(time, hazard1[, "time"])
  haz1 <- hazard1[idx1, "hazard"]
  if(missing(starttime)){
    return(haz1) }
  
  else {
    starttime <- as.character(substitute(starttime))
    ini_time <- data[, starttime]
    X <- rep(1, nrow(data))
    cox_out <- survival::coxph(survival::Surv(ini_time, time, status) ~ X)
    cox_out$coefficients["X"] <- 0
    haz <- predict(cox_out, newdata = data.frame(ini_time, time,  0, 1),
                   type = "expected")
    return(haz)
  }
}
\end{verbatim}

\end{document}